\RequirePackage{fix-cm}
\documentclass[twocolumn,epjc3]{svjour3}  
\smartqed  % flush right qed marks, e.g. at end of proof
\RequirePackage{graphicx}
\RequirePackage{amsmath}
\RequirePackage{amscd}
\RequirePackage{amssymb}
\RequirePackage{mathrsfs}
\RequirePackage{latexsym}
\RequirePackage{dsfont}
\RequirePackage{color}

\newcommand{\Eq}[1]{Eq.~(\ref{#1})}

\newcommand{\Fig}[1]{Fig.~\ref{#1}}

\newcommand{\order}[1]{\mathcal{O}\left(#1\right)}

\journalname{Eur. Phys. J. C}
\begin{document}

\title{
  Leading-order hadronic contributions to the electron and tau anomalous magnetic moments
  %\thanksref{t1}
}
%%% \subtitle{Do you have a subtitle?\\ If so, write it here}


\author{
Florian Burger \thanksref{e1,addr1}
\and
Karl Jansen  \thanksref{e2,addr2} 
\and
Marcus Petschlies \thanksref{e3,addr3,addr4} 
\and
Grit Pientka \thanksref{e4,addr1} 
}

\thankstext{e1}{e-mail: florian.burger@physik.hu-berlin.de}
\thankstext{e2}{e-mail: karl.jansen@desy.de}
\thankstext{e3}{e-mail: marcus.petschlies@hiskp.uni-bonn.de}
\thankstext{e4}{e-mail: grit.hotzel@physik.hu-berlin.de}


\institute{
Humboldt-Universit\"at zu Berlin, Institut f\"ur Physik, Newtonstr. 15, D-12489 Berlin, Germany \label{addr1}
\and
NIC, DESY, Platanenallee 6, D-15738 Zeuthen, Germany \label{addr2}
\and
The Cyprus Institute, P.O.Box 27456, 1645 Nicosia, Cyprus \label{addr3}
\and
Rheinische Friedrich-Wilhelms-Universit\"at Bonn, Institut f\"ur Strahlen- und Kernphysik, Nu{\ss}allee 14-16, D-53115 Bonn, Germany \label{addr4}
%%%  \emph{Present Address:} if needed\label{addr3}
}

\date{Received: date / Accepted: date}

\maketitle

\begin{abstract}
The leading hadronic contributions to the anomalous magnetic moments of the electron and the $\tau$-lepton
are determined by a four-flavour lattice QCD computation with twisted mass fermions.
The results presented are based on the quark-connected contribution to
the hadronic vacuum polarization function.
The continuum limit is taken and systematic uncertainties are quantified.
Full agreement with results obtained by phenomenological analyses is found.

%%% 12.38.Gc Lattice QCD calculations
%%% 12.15.Lk Radiative corrections, electroweak
\end{abstract}

\section{Introduction}
The standard model of particle physics (SM) contains three charged leptons $l$, mainly differing in mass, the electron, the muon, and the
$\tau$-lepton with $m_e : m_\mu : m_\tau \approx 1 : 207 : 3477$~\cite{Agashe:2014kda}.
Their magnetic moments, in particular their so-called anomalous magnetic moments, $a_l=(g-2)_{l}/2$, control their behaviour in an external
magnetic field.

Being the lepton with the smallest mass, the electron is stable. This leads to
the electron magnetic moment being one of the most precisely determined
quantities in nature. 
The difference between the latest experimental~\cite{Hanneke:2008tm} and SM
values~\cite{Aoyama:2012wk,Aoyama:2014sxa} is of $\order{10^{-12}}$
  or approximately
1.3 standard deviations, c.f.~\cite{Giudice:2012ms} and references therein,

\begin{eqnarray*}
  a_e^\mathrm{Exp} &=& 115\,965\,218\,07.3\,(2.8) \times 10^{−13} \\
  a_e^\mathrm{SM}  &=& 115\,965\,218\,17.8\,(7.6) \times 10^{-13}\\
  a_e^\mathrm{Exp} - a_e^\mathrm{SM} &=& -10.5\,(8.1)\times 10^{-13}\,.
\end{eqnarray*}

This constitutes one of the
cornerstone results for quantum field theories to be recognised as the 
correct mechanism for describing particle interactions. The very good 
agreement of the electron magnetic moment between experiment and 
SM calculations is not matched by the muon anomalous magnetic moment. 
In fact, 
here a two to four sigma discrepancy is observed, see e.g.~\cite{Brambilla:2014jmp}.                                   
%Since 
One reason for the observed discrepancy could be that the magnetic moment of the muon receives larger 
non - perturbative contributions than the one of the electron. On the other hand, it is supposed to be also more sensitive to beyond the SM physics,
since for a
large class of theories new physics contributions are expected to be proportional to the squared
lepton mass. Thus it is a 
prime candidate for detecting physics beyond the SM.
Due to the large mass of the $\tau$-lepton, it would be the optimal
lepton for finding new physics. However, because its lifetime is very short ($\order{10^{-13}}$s)
there currently only exist bounds on its anomalous magnetic moment from indirect measurements~\cite{Abdallah:2003xd}.

The QED~\cite{Aoyama:2012wj,Aoyama:2012wk} and the electroweak contributions~\cite{Czarnecki:1995wq,Czarnecki:1995sz} to the lepton
anomalous magnetic moments have been computed in perturbation theory to impressive five and two loops, respectively.
The main 
uncertainties remaining in the theoretical determinations of the anomalous magnetic moments originate thus from the leading-order (LO)
hadronic contributions. Since they are particularly sensitive to those virtual photon
momenta that are of $\order{m_l^2}$,
these contributions are 
inherently non-perturbative and not accessible to perturbation theory. 
In order to have a prediction of the anomalous magnetic moments from the 
SM alone, a non-perturbative method needs to be employed and the
only such approach we presently know,
which eventually allows us to control all systematic uncertainties,
is lattice QCD (LQCD) which we use here.

As highlighted in \cite{Giudice:2012ms}, the uncertainty in the comparison between the experimental and the SM value for the electron
anomalous magnetic moment is currently dominated by the experimental uncertainty of its determination and the value for $\alpha_{\rm QED}$ from
atomic
physics experiments with rubidium atoms which both are to be reduced in the future.
Recently, the Harvard group has announced to be working on a more
accurate determination of the electron as well as the positron $(g-2)$~\cite{Hoogerheide:2014mna}. According to
Ref.~\cite{Giudice:2012ms}, uncertainties in the sub-$10^{-13}$ region might be expected which
would clearly provide the opportunity to also detect new physics contributions in the anomalous magnetic moment of the electron and thus to
cross-check the muon discrepancy. In
this situation it will again be of utmost importance to know the hadronic contributions as precisely as possible.

Furthermore, even for the $\tau$-lepton, Ref.~\cite{Pich:2013lsa} lists several proposals for the first actual measurement of its anomalous magnetic moment, e.g.~\cite{Fael:2013ij}. A first successful measurement in this direction has been reported in \cite{Oberhof:2015hea}. As we will show in the following, compared to the case of the muon it will be much easier to obtain a value for the LO QCD
contribution to $a_\tau$ from
LQCD with the required precision to detect new physics and it will probably not take very long before the QCD contribution entering the official SM
result will
be provided by LQCD.

As mentioned before, the hadronic LO contributions to the
anomalous magnetic moments of the three SM leptons, 
$a_l^{\rm hvp}$, strongly depend on the values of their masses. 
Since the magnitude of the lepton masses
spans four orders of magnitude,
the corresponding contributions to the anomalous magnetic moments differ substantially and probe very 
different energy regions, see also the discussion of Fig.~\ref{fig:saturation}
in Sect.~\ref{sec:e}.

In this article, we present the results of our four-flavour computations of the quark-connected, LO hadronic vacuum polarisation contributions to the
electron and
$\tau$-lepton anomalous magnetic moments obtained from the (maximally) 
twisted mass formulation of LQCD. The muon case has already been covered in~\cite{Burger:2013jya}.
The important feature of
the present calculation is that we adopt exactly the same strategy as for the 
muon~\cite{Burger:2013jya} including the same chiral and continuum extrapolations.
Thus, the results presented here are not only interesting in themselves,
but also serve as an important cross-check for our treatment of the hadronic vacuum polarisation
function. 
  The consistent picture we obtain for all three standard model leptons then reassures
the validity of the analysis approach.

%%% \textcolor{red}{I AM SURE THIS IS WHAT WE WANT TO SAY; IS THIS TOO STRONG IN VIEW OF REFEREE 1?}
%
Additionally to the systematic uncertainties investigated in our previous paper, we quantify the light-quark disconnected contributions on
one of our $N_f=2+1+1$ ensembles
  in order to arrive at a rough estimate of their systematic effect on our estimates for the hadronic LO anomalous magnetic moments.
A full quantitative constraint on the quark-disconnected contribution, however, is beyond the scope of this work.

Another very important feature is that incorporating the complete first two generations of quarks enables us to directly and unambiguously
compare our results with the values obtained from phenomenological analyses 
relying on experimental data and a dispersion relation. We note that the contributions 
from third-generation quarks can be neglected, since they are smaller than the current theoretical accuracy, as can be inferred e.g.~from the data
tables of Ref.~\cite{Benayoun:2012wc}. Recently, the bottom quark contribution to $a_{\mu}^{\rm hvp}$ has been explicitly computed on the
lattice~\cite{Colquhoun:2014ica} confirming it to be one order of magnitude smaller than the current uncertainty of the phenomenological
determinations of $a_{\mu}^{\rm hvp}$.

Additionally to the $N_f=2+1+1$ flavour ensembles~\cite{Baron:2010bv,Baron:2010th} at unphysically large pion masses studied
in~\cite{Burger:2013jya}, we computed the dominant light quark contributions to the anomalous magnetic moments on a $N_f=2$ flavour ensemble directly
at the physical point~\cite{Abdel-Rehim:2013yaa,Abdel-Rehim:2014nka}. This allows us to test 
the chiral extrapolations performed when using the
reparametrisation introduced in~\cite{Feng:2011zk,Renner:2012fa}.
  Since we currently only have one ensemble at the physical point with one lattice spacing
  and we neglect the small influence of the strange and charm sea quarks when comparing the results from $N_f = 2$ 
  and $N_f = 2+1+1$ ensembles, a final conclusion is precluded at this point.
  The significance of this comparison is based on the empirically observed
  weak dependence on the lattice spacing of the light quark contribution as well as the marginal sea quark effects from strange and charm
  on the latter.

The next section comprises a short repetition of the most important equations needed to follow the discussion of the results for the LO
hadronic vacuum polarisation contributions to the anomalous magnetic moments of the electron in Sect.~\ref{sec:e} and the $\tau$-lepton in
Sect.~\ref{sec:tau}. In Sect.~\ref{sec:conc} we summarise our results and draw our conclusions.

\section{Computation of $a_{l}^{\rm hvp}$}
The LO hadronic contribution to the lepton anomalous magnetic moments, $a_{l}^{\mathrm{hvp}}$, can be directly computed in Euclidean
space-time according to~\cite{Lautrup:1971jf,Blum:2002ii}
\begin{equation}
a_{l}^{\mathrm{hvp}} = \alpha^2 \int_0^{\infty} \frac{d Q^2 }{Q^2} w\left( \frac{Q^2}{m_{l}^2}\right) \Pi_{\mathrm{R}}(Q^2) \; ,
\label{eq:amudef}
\end{equation}
where $\alpha$ is the fine structure constant, $Q^2$ the Euclidean momentum, $m_l$ the lepton mass, and
$\Pi_{\mathrm{R}}(Q^2)$ the
renormalised hadronic vacuum polarisation function,
$$\Pi_{\mathrm{R}}(Q^2)= \Pi(Q^2)- \Pi(0) \;. $$
 It is obtained from the hadronic vacuum polarisation tensor
\begin{eqnarray}
   \Pi_{\mu \nu}(Q) &=& \int d^4 x \,e^{iQ\cdot(x-y)} \langle J_{\mu}(x) J_{\nu}(y)\rangle \nonumber\\
   &=& \left(Q_{\mu} Q_{\nu} - Q^2 \delta_{\mu \nu}\right) \Pi(Q^2)\; ,
\label{eq:vptensor}
\end{eqnarray}
which is transverse because of the conservation of the electromagnetic current 
\begin{eqnarray}
 J_{\mu}(x) &=& \frac{2}{3} \bar{u}(x)\, \gamma_{\mu} u(x) - \frac{1}{3}\, \bar{d}(x) \gamma_{\mu} d(x)  \nonumber\\
 &&\quad + \frac{2}{3} \,\bar{c}(x) \gamma_{\mu} c(x) - \frac{1}{3} \,\bar{s}(x) \gamma_{\mu} s(x) \;.
\end{eqnarray}
Here u stands for the up quark, d for the down quark, c denotes the charm quark, and s the strange quark. 
Eq.~(\ref{eq:vptensor}) shows that $\Pi_{\mu \nu}(Q)$ results from the Fourier transformation
of the correlator of two such currents. Taking up and down quarks together, since
they are mass-degenerate in our setup, we decompose the quark-connected part of the hadronic vacuum polarisation tensor according
to 
\begin{equation}
  \Pi_{\mu \nu}(Q) = \Pi^{\rm ud}_{\mu \nu}(Q) + \Pi^{\rm s}_{\mu \nu}(Q) + \Pi^{\rm c}_{\mu \nu}(Q) \;.
  \label{eq:pimunu_flavor_decomposition}
\end{equation}

In our lattice calculation this decomposition into flavour contributions is particularly straightforward, because
for all quark flavours we use the one-point-split vector currents, which are conserved at non-zero lattice spacing and 
thus do not require further multiplicative or additive normalisation.
From eq. (\ref{eq:pimunu_flavor_decomposition}),
we can stepwise add the flavour contributions which will be done in the sections below.

The standard integral definition in Eq.~(\ref{eq:amudef}) results in a strong non-linear pion mass dependence, in
particular for the light quark contribution to $a_l^{\mathrm{hvp}}$. 
This behaviour originates from
the introduction of the lepton mass $m_l$ as an external scale, which is not related
to the lattice parameters and in particular does not have an inherent value in lattice units.
Employing Eq. (\ref{eq:amudef})  in the lattice calculation requires the input
of the dimensionless combination
$a\cdot m_l$ and this renders the initially dimensionless quantity $a_l^{\mathrm{hvp}}$ effectively dependent
on the lattice scale setting. In view of this external scale problem, Refs.~\cite{Feng:2011zk,Renner:2012fa} 
proposed a modified definition of a new family of observables
\begin{equation}
  a_{\bar{l}}^{\mathrm{hvp}} = \alpha^2 \int_0^{\infty} \frac{d Q^2 }{Q^2} w\left( \frac{Q^2}{H^2}
\frac{H_{\mathrm{phys}}^2}{m_{l}^2}\right) \Pi_{\mathrm{R}}(Q^2)\,.
\label{eq:redef}
\end{equation}
$H$ denotes some hadronic scale determined on the lattice at unphysically high pion masses, which fulfills the constraint, that
$H\left( m_{\mathrm{PS}} \right) \to H_\mathrm{phys}$ as $m_\mathrm{PS} \to m_\pi$. For each choice
of $H$ we thus obtain a correspondingly modified, $m_\mathrm{PS}$ dependent lepton mass on the lattice 
$m_{\bar{l}}\left( m_{\mathrm{PS}} \right) = m_l \cdot H \left( m_\mathrm{PS} \right) / H_\mathrm{phys}$.

  Our choice for the hadronic scale $H$ is the mass $m_V$ of the lowest-lying state in the
  light vector meson channel, i.e. the $\rho$-meson state.
  This choice uniquely fixes the pion mass dependence of the lepton mass $m_{\bar{l}}\left( m_\mathrm{PS} \right)$
  and is subsequently used for all single-flavour contributions to the vacuum polarization function.

$H=H_{\rm phys}=1$ reproduces the standard
definition in Eq.~(\ref{eq:amudef}). Up to lattice artefacts, the standard definition is also recovered at the physical 
value of the pion mass when the ratio $H/H_{\rm phys}$ becomes one
\begin{equation*}
  \lim\limits_{m_\mathrm{PS} \to m_\pi} \,a_{\bar{l}}^\mathrm{hvp}\left( m_\mathrm{PS} \right) 
    = a_l^\mathrm{hvp}\left( m_\pi \right)\,.
\end{equation*}
Henceforth we always use the definition in Eq. (\ref{eq:redef}) with $H = m_V$ and drop the bar on the label for the lepton.

The weight function $w$ is known from QED perturbation theory as
\begin{equation}
   w(r) = \frac{64}{r^2 ( 1 + \sqrt{ 1 + 4/r } )^4 \sqrt{ 1 + 4/r }}\,.
   \label{eq:weightfun}
 \end{equation}
 It has a pronounced peak at $r_\mathrm{peak} = Q^2_{\mathrm{peak}} / m_{l}^2 = \sqrt{5} - 2$.
 As an illustrating example the corresponding peak locations for the electron, muon and tau are shown
 %%% by the arrows pointing on the abscissa
 by the labels on the upper $x$-axis
 in the upper plot of Fig. \ref{fig:saturation}
 for ensemble D30.48 in Table \ref{tab:ensemble_table} below.
% \begin{equation}
%  Q^2_{\rm peak}=(\sqrt{5}-2) \frac{H^2}{H_{\rm phys}^2} m_l^2 \; .
% \end{equation}

For a thorough description of the lattice calculation and a proof of automatic $\mathcal{O}(a)$ improvement of the vacuum polarisation function we
refer to~\cite{Burger:2013jya} and~\cite{Burger:2014ada}, respectively.
In order to discuss systematic uncertainties later on, we briefly summarise our method of fitting the hadronic vacuum polarisation function here.                 

First, the lowest lying vector meson masses, $m_i$, and decay constants, $f_i$, are determined from the time
dependence of the two-point function of
the light, strange, and charm point-split vector current, individually, at zero spatial momentum.
Then $\Pi(Q^2)$  determined  in the momentum range between $0$ and $Q^2_{\rm max}$ is split  into a low-momentum
part for $ 0 \le Q^2 \le Q^2_{\rm
match}$ and a high-momentum one for $ Q^2_{\rm match} < Q^2 \le Q^2_{\rm max}$ and is fitted separately for each flavour and each
ensemble. The low-momentum fit function is given by
\begin{equation}
  \Pi_{\mathrm{low}}(Q^2) = \sum_{i=1}^M \frac{f^2_i}{m^2_i + Q^2} + \sum_{j=0}^{N-1} a_j (Q^2)^{j} \; ,
\label{eq:pilow}
\end{equation}
and the high-momentum piece is parametrised as follows
\begin{equation}
  \Pi_{\mathrm{high}}(Q^2) = \log(Q^2) \sum_{k=0}^{B-1} b_k (Q^2)^{k}  + \sum_{l=0}^{C-1} c_l (Q^2)^{l} \; . 
\label{eq:pihigh}
\end{equation}
They are combined according to
\begin{eqnarray}
 \Pi(Q^2) &=& (1- \Theta(Q^2-Q^2_{\rm match}))\Pi_{\mathrm{low}}(Q^2) \nonumber\\
 &&\quad + \Theta(Q^2-Q^2_{\rm match}) \Pi_{\mathrm{high}}(Q^2) \; ,
\label{eq:pilowandhigh}
\end{eqnarray}
where $\Theta(x)$ is the Heaviside function.

This defines our so-called MNBC fit function.  Our standard fit for the light and strange quark contributions is M1N2B4C1 which means $M=1$, $N=2$,
$B=4$, and $C=1$ in Eqs.~(\ref{eq:pilow}) and (\ref{eq:pihigh})
above. As value of $Q^2_{\rm match}$ in the Heaviside functions in Eq.~(\ref{eq:pilowandhigh}) we have chosen $2\,{\rm GeV}^2$. We have checked
that varying the value of $Q^2_{\rm match}$ between $1\,{\rm
 GeV}^2$ and $3\,{\rm GeV}^2$ does not lead to observable differences as long as the transition between the low- and the high-momentum part of the
fit is smooth. For the upper integration limit we use $Q^2_{\rm max} = 100\,{\rm GeV}^2$, since the integrals are saturated there as can be seen in
Fig.~\ref{fig:saturation} below. 

  For each ensemble and flavour we perform a fit for $\Pi$ as given in Eq.~(\ref{eq:pilowandhigh}). From this fit we obtain the corresponding
  $\Pi(0)$ and thus the subtracted polarisation function. The latter is integrated using Eq.~(\ref{eq:redef}) and
  the contributions from individual quark flavours are summed including the appropriate charge factors 
  $ e_\mathrm{u}^2 + e_\mathrm{d}^2$, $e_\mathrm{s}^2$, and $e_\mathrm{c}^2$.
  This results in an estimate for the hadronic leading-order lepton anomalous magnetic moment for each gauge field ensemble, 
  which depends on the lattice spacing, the pion mass,
  and the lattice size, 
  $ a_{l}^{\mathrm{hvp}} = a_{l}^{\mathrm{hvp}}\left( a, m_{\mathrm{PS}}^2, L\right) $. 
  As final step we perform a combined extrapolation to the continuum and to physical quark masses. For this extrapolation some
  dependencies turn out to be negligible. The strange and charm quark mass in the sea and valence
  quark action have been tuned for each ensemble, so we do not need to consider these dependencies explicitly in $a_{l}^{\mathrm{hvp}}$.
  Moreover, the detailed discussion of the lattice data in the following sections will show
  that we only find significant lattice artefacts in the strange and charm quark contribution to $a_{l}^{\mathrm{hvp}}$ and we will
  use an appropriately adapted fit ansatz. The dependence on the finite volume is discussed in detail as part of the 
  systematic error analyses in sections \ref{sec:FSE_ae} and \ref{sec:FSE_atau} and thus not part of the extrapolation described above.
  %%% In general, we neglect finite size effects in our extrapolations. This will later be discussed
  %%% in detail as part of the systematic error analysis. 

% In~\cite{Aubin:2012me,Golterman:2013vca} the usage of Pad\'e approximants has been advocated in order to avoid systematic errors
% related to the choice of fit function. We have analysed these fit types in~\cite{Burger:2014lna} and found agreement between the values of the poles
% and the results for $a_{\rm e}^{\rm hvp}$ between our MNBC fits and Pad\'e fits provided the same number of parameters have been
% used. A more elaborate study for the case of the muon has been performed by the RBC-UKQCD collaboration in~\cite{Marinkovic:2015zaa}. There it was
% shown
% that it is not clear which
% type of Pad\'e fit and which upper fit limit results in a correct answer.

  In~\cite{Aubin:2012me,Golterman:2013vca} the usage of Pad\'e approximants has been advocated for fits in the small momentum region.
  The Pad\'e fit functions are formally identical to the $MN$ series of fits. We analysed the Pad\'e approximants for
  our data in \cite{Burger:2014lna}. We found agreement for the location of the poles provided the same number of fit parameters were
  used in both cases (See also the more elaborate study for the case of the muon performed by the RBC-UKQCD collaboration in~\cite{Marinkovic:2015zaa}.).
  We can thus employ the same procedure as for the muon and show that it produces results compatible with phenomenological determinations
  for both the electron and the $\tau$-lepton without any modification.

%%% \textcolor{red}{This is one reason why} we stick with our standard procedure
%%% until data with significantly reduced statistical uncertainties on ensembles of much larger size become available.
%%% \textcolor{red}{The main reason for employing the same procedure as for the muon is, of course, to show that it produces results compatible with phenomenological determinations for both the electron and the $\tau$-lepton without any modification.}

Our analysis has been performed on the same set of gauge field configurations~\cite{Baron:2010bv,Baron:2010th}
as have been used in our previous work \cite{Burger:2013jya}.
A detailed list of the lattice parameters can be found in table
  \ref{tab:ensemble_table} below.

\begin{table}[htb]
  \begin{center}
    \begin{tabular}{|c | c c c c|}
      \hline
      & & & &\vspace{-0.40cm} \\
      Ensemble  &$a[{\rm fm}]$ & $m_{PS}$[MeV] &$ L$[fm] & $m_{PS}\cdot L$\\
      & & & &\vspace{-0.40cm} \\
      \hline \hline
      & & & &\vspace{-0.40cm} \\
      D15.48   & $0.061$ & 227 & 2.9  & $3.3$ \\
      D30.48   & $0.061$ & 318 & 2.9  & $4.7$ \\
      D45.32sc & $0.061$ & 387 & 1.9  & $3.8$ \\
      % & & & & \vspace{-0.40cm} \\
%
      \hline
      & & & & \vspace{-0.40cm} \\
      B25.32t& $0.078$ & 274 &  2.5 & $3.5$ \\
      B35.32 & $0.078$ & 319 &  2.5 & $4.0$ \\
      B35.48 & $0.078$ & 314 &  3.7 & $5.9$ \\
      B55.32 & $0.078$ & 393 &  2.5 & $5.0$ \\
      B75.32 & $0.078$ & 456 &  2.5 & $5.8$ \\
      B85.24 & $0.078$ & 491 &  1.9 & $4.7$ \\
      & & & &\vspace{-0.40cm} \\
      \hline
      & & & &\vspace{-0.40cm} \\
      A30.32 & $0.086$ & 283 & 2.8 & $4.0$ \\
      A40.32 & $0.086$ & 323 & 2.8 & $4.6$ \\
      A50.32 & $0.086$ & 361 & 2.8 & $5.1$ \\
      \hline
      & & & &\vspace{-0.40cm} \\
      cA2.09.48 & $0.091$ & $135$ & 4.4 & $3.0$ \\
      \hline
    \end{tabular}
    \caption{ Parameters of the $N_f = 2+1+1$ flavour gauge field configurations that
      have been analysed in this work. $a$ denotes the lattice spacing (cf. \protect\cite{Baron:2011sf}),
      $m_{PS}$ the value of the light pseudoscalar meson mass (cf. \protect\cite{Baron:2010bv})
      and $L$ the spatial extent of the lattices. 
      The right-most column gives the value for $m_{PS}\cdot L$. The ensemble in the last line has $N_f =2$ and physical pion mass.
      It is described in Refs. \protect\cite{Abdel-Rehim:2013yaa,Abdel-Rehim:2014nka,Abdel-Rehim:2015pwa}.
        The ensemble name in the first column gives the bare quark mass in lattice units as the first pair
        of digits times $10^{-4}$ and the spatial lattice size $L/a$ as the second pair of digits.
    }
    \label{tab:ensemble_table}
  \end{center}
\end{table}

\section{The electron $(g-2)$}
\label{sec:e}
\begin{figure}[htbp]
 \centering
\includegraphics[width=0.45\textwidth]{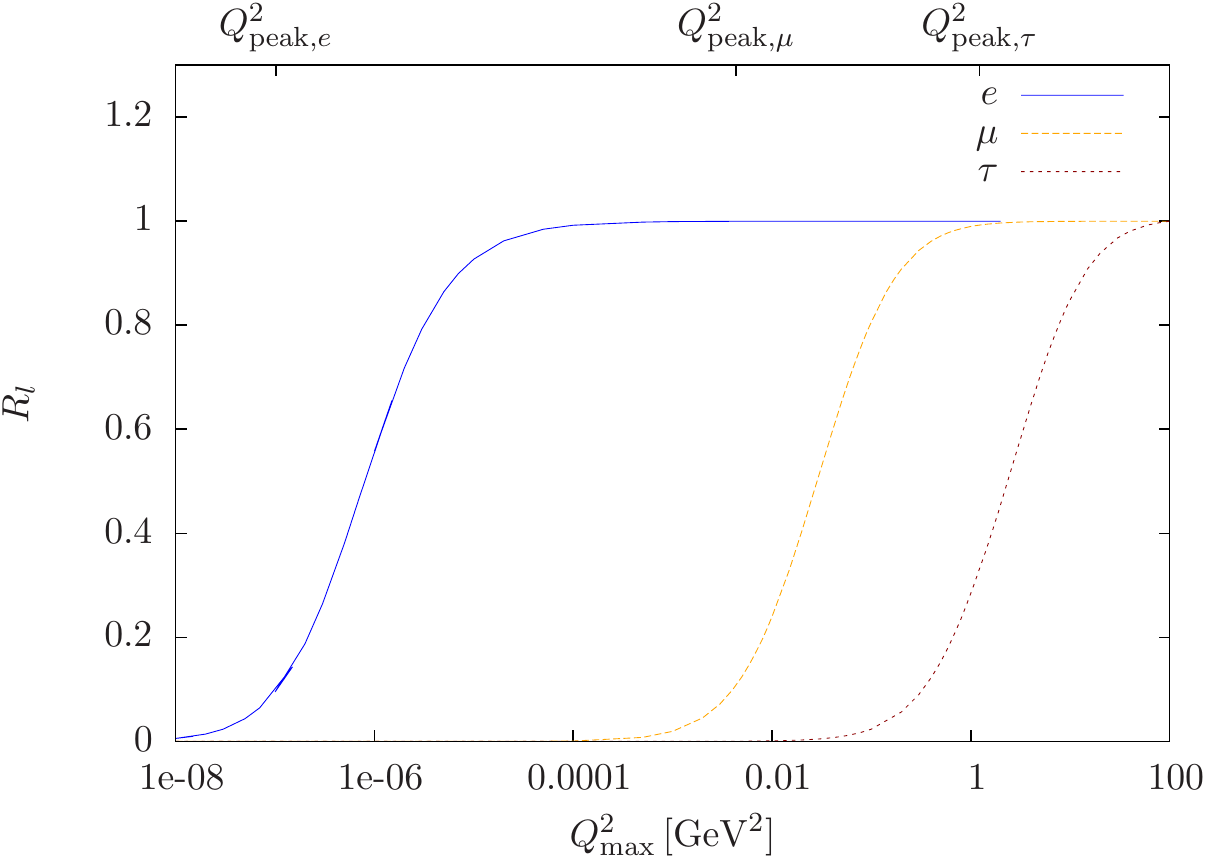}
\includegraphics[width=0.45\textwidth]{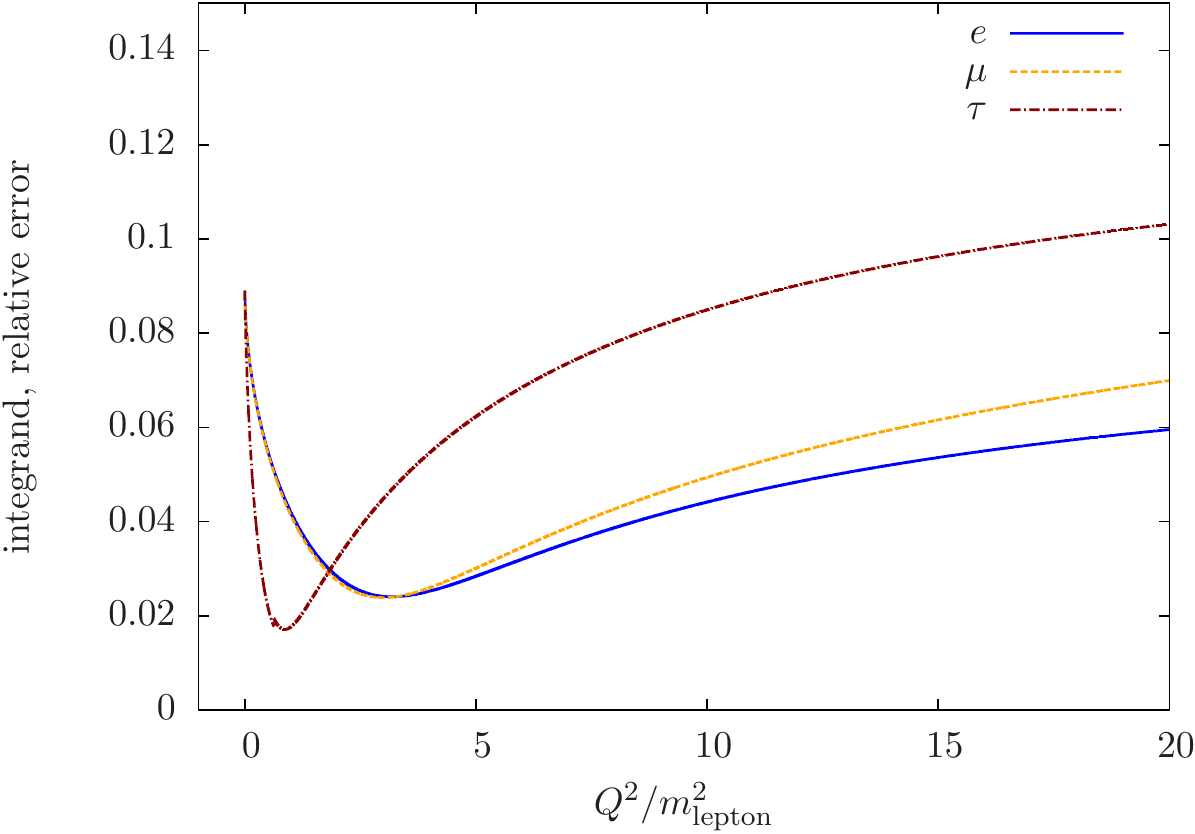}
\hfill
\caption{Upper plot: Comparison of the dependence on the upper integration bound in Eq.~(\ref{eq:redef}) of the four-flavour lepton anomalous magnetic
moments.
The blue curve represents the ratio defined in Eq.~(\ref{eq:ratio}) for the electron, the orange one for the muon, and the dark red one for the tau.
$Q^2_{{\rm peak}, l}$ denotes the momentum value where the kernel function in Eq.~(\ref{eq:redef}) attains its maximum. Lower plot:
Comparison of the dependence of the relative statistical uncertainties of the integrands in Eq.~(\ref{eq:redef}) on the squared momenta scaled by the
lepton masses. The plots are based on data for the D30.48 ensemble featuring $a = 0.061\,\mathrm{fm},\,m_{\mathrm{PS}} = 318\,\mathrm{MeV}$, and $L=2.9\,\mathrm{fm}$.  }
\label{fig:saturation}
\end{figure}

The LO hadronic contribution to
the electron anomalous magnetic moment $a_{\rm e}$
is dominated by momenta below $10^{-4}\,{\rm GeV}^2$. To a good approximation it can even be determined from the slope
of the vacuum polarisation at zero momentum $a_{\rm e} \propto d \Pi / dQ^2 (Q^2=0)$.
Therefore, we only use the low-momentum part, $\Pi_{\mathrm{low}}(Q^2)$, of the hadronic vacuum
polarisation function Eq.~(\ref{eq:pilow}). The saturation of the integral for one of our ensembles, namely B55.32 featuring $m_{\rm PS} \approx
390\, {\rm MeV}$, $a\approx 0.08\,{\rm fm}$ and $L = 2.5 \, {\rm fm}$, is shown in the upper plot of Fig.~\ref{fig:saturation} for
all three leptons by plotting
\begin{equation}
 R_l (Q^2_{\rm max}) = \frac{a_l^{\rm hvp}(Q^2_{\rm max})}{a_l^{\rm hvp}(100\,{\rm GeV}^2)} \; ,
\label{eq:ratio}
\end{equation}
where $a_l^{\rm hvp}(Q^2_{\rm max})$ is the LO hadronic contribution to the lepton anomalous magnetic moment integrated up to $Q^2_{\rm max}$.
This plot also implies that for the electron we have to rely 
mostly on the extrapolation of our vacuum polarisation data to the small momentum region.
Although saturating well beyond momenta of $\mathcal{O}\left( 1\,\mathrm{GeV}^2 \right)$, also for the $\tau$, the renormalised vacuum
polarisation function requires the subtraction of $\Pi(0)$, which is determined from the same extrapolation to the small momentum region.

Despite the different masses of the
leptons, the lower plot of Fig.~\ref{fig:saturation} shows that the relative statistical uncertainties are
of the same orders of magnitude for all
three leptons and display a universal dependence on $Q^2/m_l^2$.

\subsection{Contribution from up and down quarks}
\begin{figure}[!htbp]
 \centering
\includegraphics[width=0.48\textwidth]{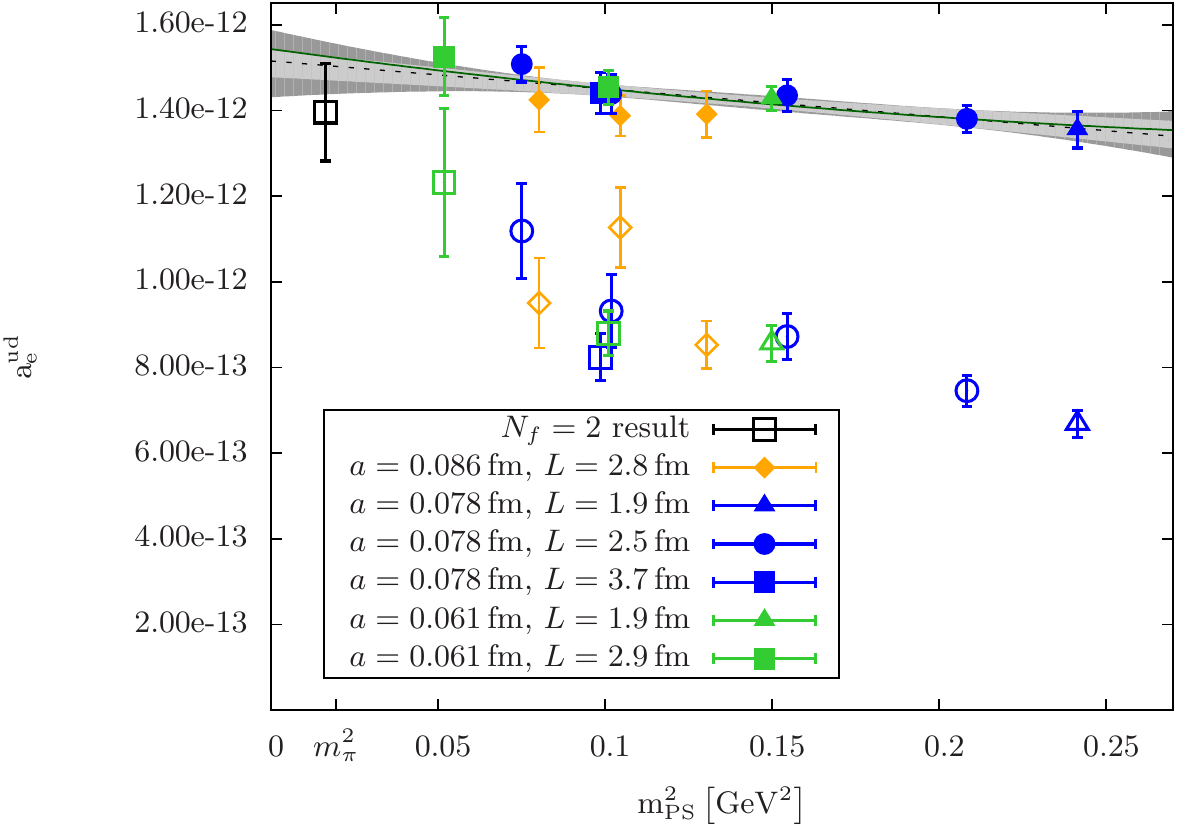}
\includegraphics[width=0.48\textwidth]{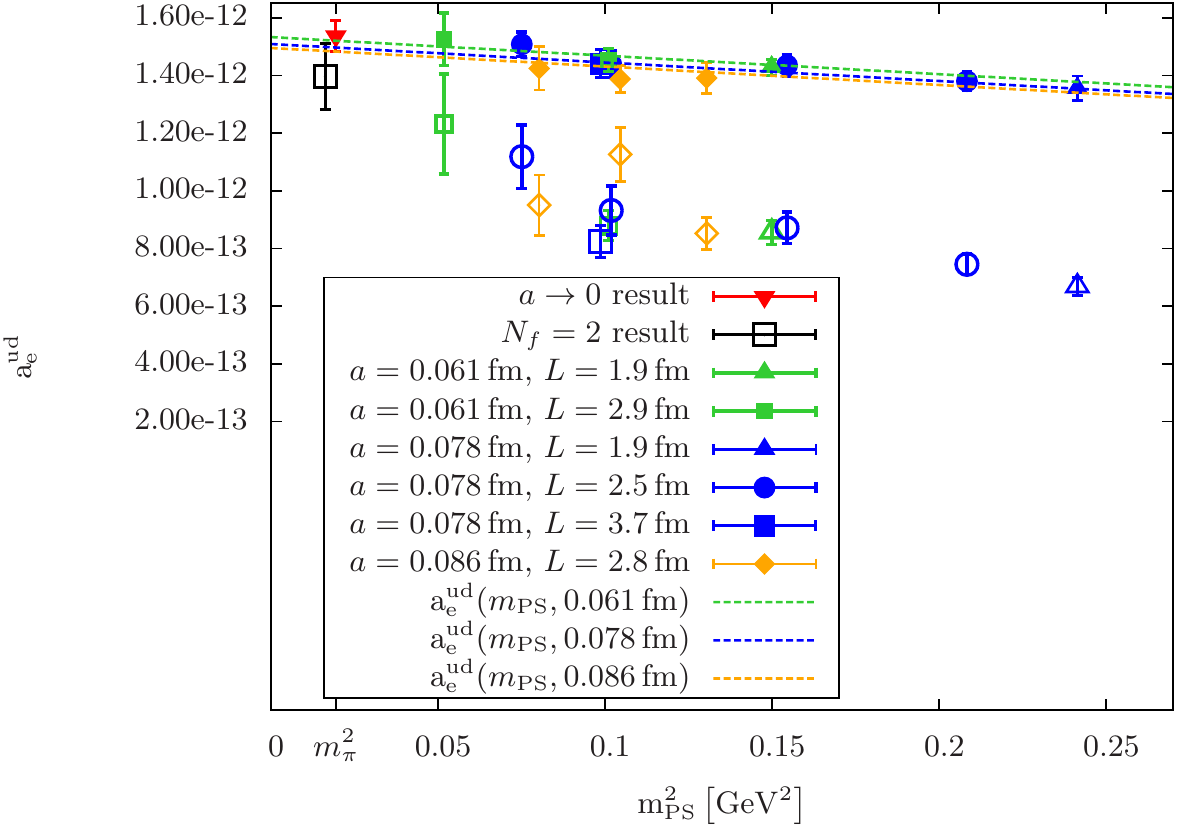}
\caption{
  Upper plot: light-quark contribution to $a_{\rm e}^{\rm hvp}$ with filled symbols representing points obtained with Eq.~(\ref{eq:redef}), open symbols
  refer to those obtained with  Eq.~(\ref{eq:amudef}), i.~e.~$H=1$. In particular,
  the two-flavour result at the physical point has been computed with the standard
  definition. The light grey errorband belongs to the linear fit, whereas the dark grey errorband is attached to the quadratic fit.
  Lower plot: combined chiral and continuum extrapolation of the light-quark contribution to $a^\mathrm{hlo}_\mathrm{e}$ allowing for
  lattice artefacts.
}
\label{fig:aelight_wphys}
\end{figure}

The light quark contribution is depicted in
Fig.~\ref{fig:aelight_wphys}. Here, we compare $a_{\rm e}^{ud}$ with
the result at the physical value of the pion mass obtained
with the standard definition
Eq.~(\ref{eq:amudef}) on one ensemble~\cite{Abdel-Rehim:2013yaa,Abdel-Rehim:2014nka} with only one lattice spacing.

  For the extrapolation in the upper plot of 
  Fig.~\ref{fig:aelight_wphys} we initially use $a_{\rm e}^{ud}$ from all our ensembles, i.e. all
  volumes, lattice spacings, and pion masses. As the figure shows, with the present accuracy of the data
  we do not resolve any statistically significant dependence of $a^{ud}_e$ on the lattice spacing or
  the lattice size. 
  Using the modified definition (Eq.~\ref{eq:redef}) with $H=m_V$ the ansatz
  \begin{equation}
    a_{\rm e}^{ud}\left( a, m_\mathrm{PS}^2, L \right) = A + B\,m_{\mathrm{PS}}^2
    \label{eq:chiral_extrapolation_formular_linear}
  \end{equation}
  is then already sufficient
  to describe our data. The result of this extrapolation is shown as the light grey band in Fig.~\ref{fig:aelight_wphys}.
  In principle, we are to add terms of higher order in $m_{\mathrm{PS}}^2$ for the extrapolation formula in Eq. (\ref{eq:chiral_extrapolation_formular_linear}),
  \begin{equation}
    a_{\rm e}^{ud}\left( a, m_\mathrm{PS}^2, L \right) = A + B\,m_{\mathrm{PS}}^2 + B_2\,m_{\mathrm{PS}}^4 + \ldots
    \label{eq:chiral_extrapolation_formular_higher_order}
  \end{equation}
  to account for any non-linear dependence on $m_{\mathrm{PS}}^2$.
  The dark grey band in the background in Fig. \ref{fig:aelight_wphys}
  shows the extrapolation with an additional term $B_2\,m_{\mathrm{PS}}^4$. The difference
  between this extrapolation and the previous one linear in the squared pion mass is insignificant.
  This insignificance of terms in $m_{\mathrm{PS}}^2$ beyond the linear one
  when using the improved definition in Eq. (\ref{eq:redef}) turns out to
  be a universal property of $a_{\rm l}^{f}$ for all leptons and all flavour combinations $f = ud, udsc$
  considered here and below. In addition, we check explicitly for lattice artefacts in the extrapolation by adding
  a term $C\,a^2$ to Eq. (\ref{eq:chiral_extrapolation_formular_linear}) with the result shown in the lower plot of
  Fig. \ref{fig:aelight_wphys}.

  We assume that lattice artefacts for the data at the physical point are negligible as well,
  which gives merit to the observed compatibility of the physical point result with the 
  value determined by the linear extrapolation of the data described here.

\subsection{Adding the strange and the charm quark contributions}
When incorporating the heavy, second-generation flavours, which are described by the Osterwalder-Seiler
action~\cite{Osterwalder:1977pc,Frezzotti:2004wz} and whose masses have been tuned to their physical values as
shown in~\cite{Burger:2013jya}, we take $\mathcal{O}(a^2)$ lattice artefacts into account. 
The four-flavour result for  $a_{\rm e}^{\rm hvp}$ at the physical point in the continuum limit is obtained from simultaneously extrapolating
in the pion mass, $m_{\rm
PS}$, and to zero lattice spacing $a$ using the ansatz
\begin{equation}
  a_{\rm e}^{\rm hvp}\left( m_{\rm PS}, a \right) = A + B\, m_{PS}^2 + C\,a^2\,.
\label{eq:fit}
\end{equation}
%%% which neglects finite size effects.
$A, B, C$ denote the free parameters of the fit.
  In the presence of the strange and charm contributions to $a_{\rm e}^{\rm hvp}$, the parameter 
  $C$ will also contain terms $\sim m_{c,R}^2,\,m_{s,R}^2$ from the renormalised charm and strange quark mass
  and also receives contributions from lattice artefacts possibly present in the
  light quark contribution to $a_{\rm e}^{\rm hvp}$.
  The reason for omitting a linear term in $a$ is that automatic $\mathcal{O}(a)$ improvement is retained for our definition
of the hadronic vacuum polarisation function at maximal twist~\cite{Burger:2014ada}.
As we have discussed in~\cite{Burger:2013jya}, systematic
effects from varying the heavy valence and sea quark masses within the range given there have been found to be negligible.
This is partly due to the contribution of strange and charm quark current correlators to vacuum polarisation being at least an
order of magnitude smaller than those from the light quarks.

\begin{figure}[htb]
 \centering
\includegraphics[width=0.48\textwidth]{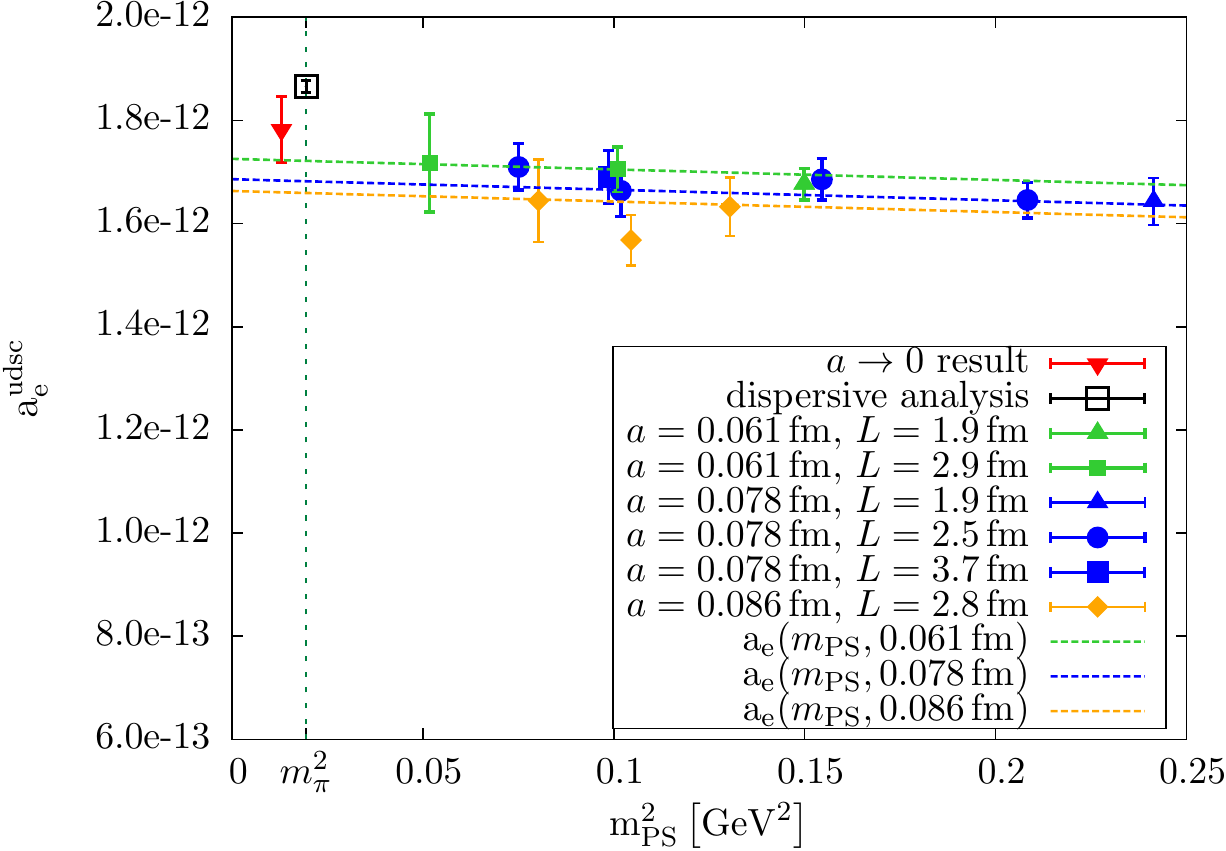}
\caption{Chiral and continuum extrapolation of the $N_f=2+1+1$ 
contribution to $a_{\rm e}^{\rm hvp}$. The inverted red triangle shows the value extrapolated 
to the continuum and to 
the physical value of the pion mass. It has been displaced to the left to
facilitate the comparison with the dispersive result in the black square~\cite{Nomura:2012sb}.}
\label{fig:ae_tot}
\end{figure}

For $a_{\rm e}^{\rm hvp}$ the corresponding fit is shown
in \Fig{fig:ae_tot}.  
  Again we use data for $a_{\rm e}^{\rm hvp}$ for all lattice spacings, pion masses, and lattice volumes in
  this extrapolation. The ansatz in Eq.~(\ref{eq:fit}) is sufficient for a good description of all our data. This is shown
  by the three dashed lines, which evaluate the fit function for the individual lattice spacings. We have checked that amending
  the fit function by higher powers in $m_\mathrm{PS}^2$ does not lead to significantly different results for the
  extrapolated value.
  Comparing the results for different lattice volumes for lattice spacing $a = 0.078\,\mathrm{fm}$ in \Fig{fig:ae_tot} 
  suggests the absence of observable finite volume effects. However, for the compilation of our complete error budget we investigate
  these effects in more detail below.

Our result with only statistical uncertainty is
\begin{equation} 
 a_{\rm e}^{\rm hvp} = 1.78\,(06)\cdot10^{-12} \;. 
\label{eq:ae_tot}
\end{equation}

\subsection{Systematic uncertainties}
In this section, we give an account of systematic uncertainties of our result for $a_{\rm e}^{\rm hvp}$ given in Eq.~(\ref{eq:ae_tot}). 
We have investigated finite size effects (FSE), the dependence of our chiral extrapolation on the incorporation of large pion masses, 
vector meson fit ranges, and the dependence of our results on different vacuum polarisation fit functions. Moreover, for one ensemble the light quark-disconnected contribution is quantified.

\subsubsection{Finite size effects}
\label{sec:FSE_ae}
As described in detail in Ref.~\cite{Burger:2013jya}, the $N_f=2+1+1$ ensembles analysed in this work feature 
$3.35 < m_\mathrm{PS}~L < 5.93$, where $L$ is the spatial extent of the lattice. 
Restricting our data to the condition $m_{\rm PS} L > 3.8$ and $m_\mathrm{PS} L > 4.5$, respectively, yields
\begin{align}
  a_{\rm e}^{\rm hvp}\left( m_\mathrm{PS} L > 3.8 \right) &= 1.77\,(07)\cdot 10^{-12} \; ,
  \label{eq:fse_a_e_mpsL_gt_3.8}\\
  a_{\rm e}^{\rm hvp}\left( m_\mathrm{PS} L > 4.5 \right) &= 1.83\,(10)\cdot 10^{-12}\; ,
  \label{eq:fse_a_e_mpsL_gt_4.5}
\end{align}
after combined chiral and continuum extrapolation. This matches the result given in Eq.~(\ref{eq:ae_tot}) and thus indicates that FSE are
negligible in our computation. This finding is supported by comparing the results of two ensembles only differing in lattice size provided in
Tab.~\ref{tab:FSE_ae}. 
%the table below.
The numbers do not change when restricting the momenta of the larger ensemble to those of the smaller one. The FSE
attributed to the lowest achievable momentum being $\frac{2\pi}{L}$ mixes with
FSE entering the choice of different fit functions. We take a conservative approach and consider these effects separately.
\begin{table}[htb]
\begin{center}
\begin{tabular}{|c c| c c|}
\hline
 & & & \vspace{-0.40cm} \\
Ensemble & $\left(\frac{L}{a}\right)^3 \times \frac{T}{a}$ & $a_{\mathrm{e, ud}}^{\mathrm{hvp}}$& $a_{\mathrm{e}}^{\mathrm{hvp}}$ \\
\hline \hline
 & & & \vspace{-0.40cm} \\
B35.32 & $32^3 \times 64$ & $1.44(05)\cdot 10^{-12}$ &   $1.66(05)\cdot 10^{-12}$ \\
B35.48 & $48^3 \times 96$ & $1.44(05) \cdot 10^{-12}$ &  $1.69(05)\cdot 10^{-12}$\\
\hline
\end{tabular}
 \caption{\label{tab:FSE_ae} Comparison of light-quark contribution to $a_{\mathrm{e}}^{\rm hvp}$ 
 and total $a_{\mathrm{e}}^{\mathrm{hvp}}$ from
 ensembles of different volumes.} 
\end{center}
\end{table}

\subsubsection{Chiral extrapolation}
We have checked the validity of the chiral extrapolation by restricting the data, comprising pion masses between $227\,{\rm MeV}$ and  $491\,{\rm MeV}$, to the condition $m_{\rm PS} < 400\,{\rm MeV}$. The value we obtain 
\begin{equation}
 a_{\rm e}^{\rm hvp} = 1.78\,(07)\cdot 10^{-12}
\end{equation}
only features a slightly larger uncertainty compared to the result in Eq.~(\ref{eq:ae_tot}). Thus, we do not assign a systematic uncertainty to
the usage of pion masses above $400\,{\rm MeV}$.

\subsubsection{Vector meson fit ranges}
Our standard computation involves the determination of the
masses and decay constants of the vector meson ground states
for the different flavours. Their values depend on the choice of fit ranges.
We have analysed different fit ranges for the two-point functions of the light, strange, and charm vector currents and propagated the uncertainties to
the values for $a_{\rm e}^{\rm hvp}$. This showed that excited state contaminations are significant only for $m_V$ and $f_V$ determined from the light
vector current-current correlator. Variations of
the standard fit ranges by $0.1\,{\rm fm}$ to the left, right and both simultaneously do not lead to any observable
differences in $a_{\rm e}^{\rm hvp}$ for the $\overline{s} \gamma_\mu s$- and the $J/\psi$ correlator. Furthermore, the heavy flavour contributions are approximately one order of magnitude smaller than the light quark contribution such that their systematic uncertainties would not noticeably impact the overall uncertainty of $a_{\rm e}^{\rm hvp}$.

\begin{figure}[htbp]
 \centering
\includegraphics[width=0.42\textwidth]{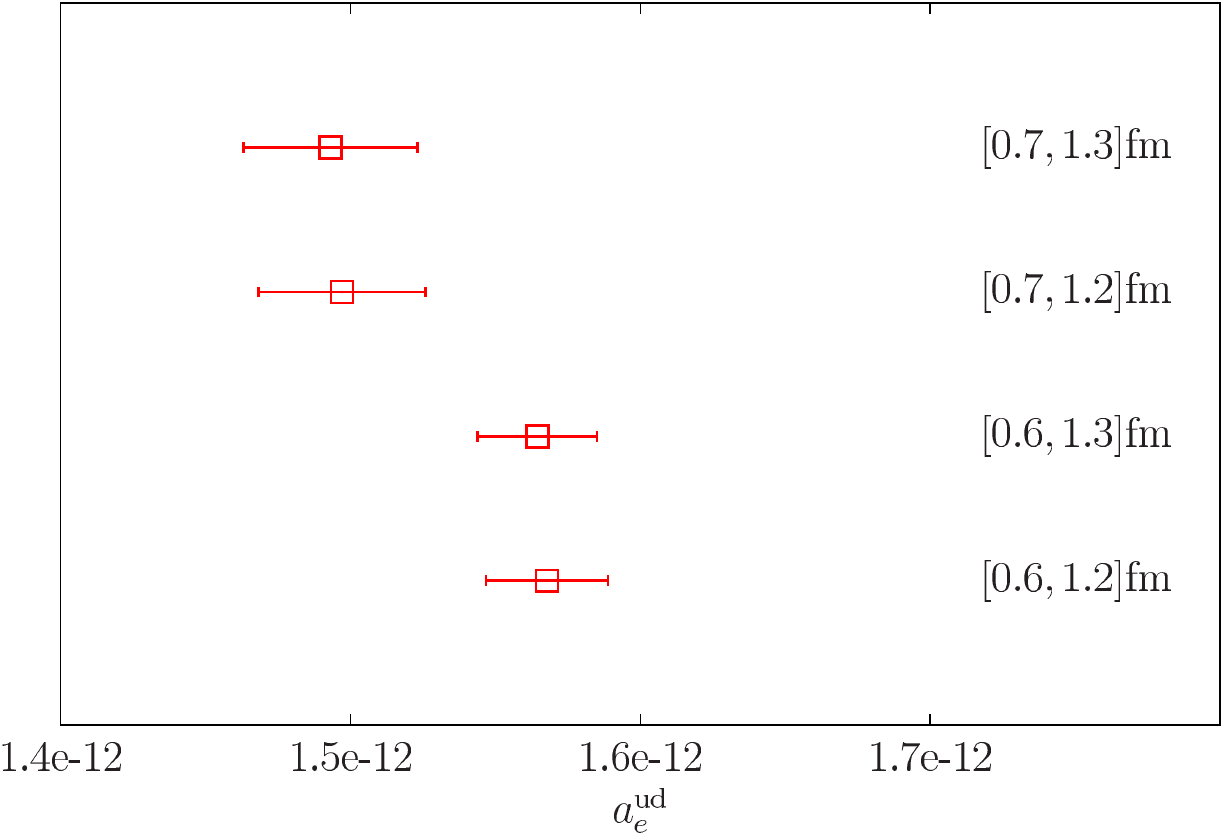}
\includegraphics[width=0.42\textwidth]{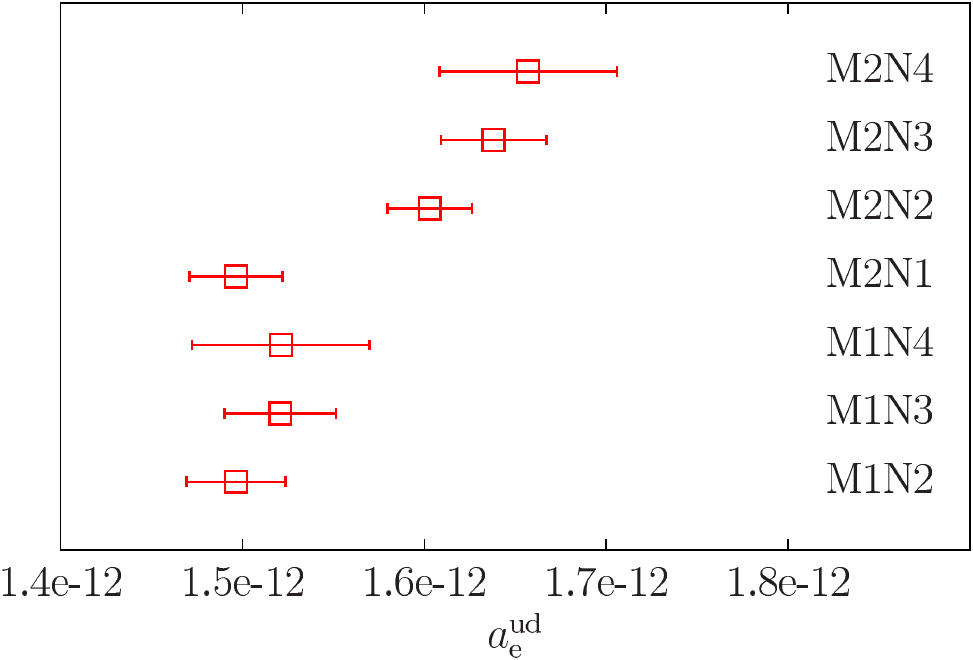}
\caption{Dependence of $a_{\rm e}^{\rm ud}$ on the fitrange of the $\rho$-correlator (upper plot) and
on values chosen for M, N in the vacuum polarisation fit function (lower plot). 
The standard $\rho$-correlator fit range is $[0.7\, {\rm fm},1.2\,{\rm fm}]$ and the standard fit function corresponds to M1N2.}
\label{fig:ae_fitrange_mnbc_light}
\end{figure}

In the upper plot of \Fig{fig:ae_fitrange_mnbc_light}, the dependence of the 
light quark contribution to the electron anomalous magnetic moment on the
fitrange for the $\rho$-correlator is plotted.
The lower limit $0.6 \,\mathrm{fm}$ is a lower bound for the
  time region, where a single-state fit describes the 2-point correlation function. The upper end at $1.3\,\mathrm{fm}$ is less
  stringent, but the signal to noise ratio for the masses deteriorates quickly and the differences in a plot like Fig. \ref{fig:ae_fitrange_mnbc_light}
  become insignificant.

Taking half the difference of the central values obtained for $[0.6\, {\rm fm},1.2\,{\rm fm}]$ and $[0.7\, {\rm fm},1.2\,{\rm fm}]$ gives a 
systematic uncertainty of
\begin{equation}
 \Delta_{V} = 0.035 \cdot 10^{-12} \; .
\end{equation}

\subsubsection{Number of terms in MN fit function}
The number of terms in the fit function Eq.~(\ref{eq:pilow}) is given by M and N. M1N2 is our standard
choice. Repeating the whole analysis with
different numbers of terms for the light quark contribution
leads to the results shown in the lower plot of Fig.~\ref{fig:ae_fitrange_mnbc_light}.
We observe that the chirally and continuum extrapolated results of fit functions involving one
and two poles are not compatible and thus we assign
a systematic error 
by taking half the difference of the central values of the 
result of the M2N3 and the M1N2 fit. 
This leads to a 
systematic uncertainty of 
\begin{equation}
 \Delta_{MN}^{\rm ud} = 0.071 \cdot 10^{-12} \; .
\end{equation}
This results in the dominant systematic uncertainty of the determination of $a_{\rm e}^{\rm hvp}$.
%%% and needs to be scrutinised further when more accurate data becomes available.
For the strange quark the systematic uncertainty from different values of M and N is
\begin{equation}
 \Delta_{MN}^{\rm s} = 0.007 \cdot 10^{-12} \;
\end{equation}
which we add to the light quark one. The differences of results from different fit functions for the charm quark contribution have turned out to be negligible such
that the total systematic error originating from employing various numbers of terms in the fit function amounts to
\begin{equation}
 \Delta_{MN} = 0.078 \cdot 10^{-12} \; .
\end{equation}

\subsubsection{Disconnected contributions}
\begin{figure}[htb]
\centering
\includegraphics[width=0.48\textwidth]{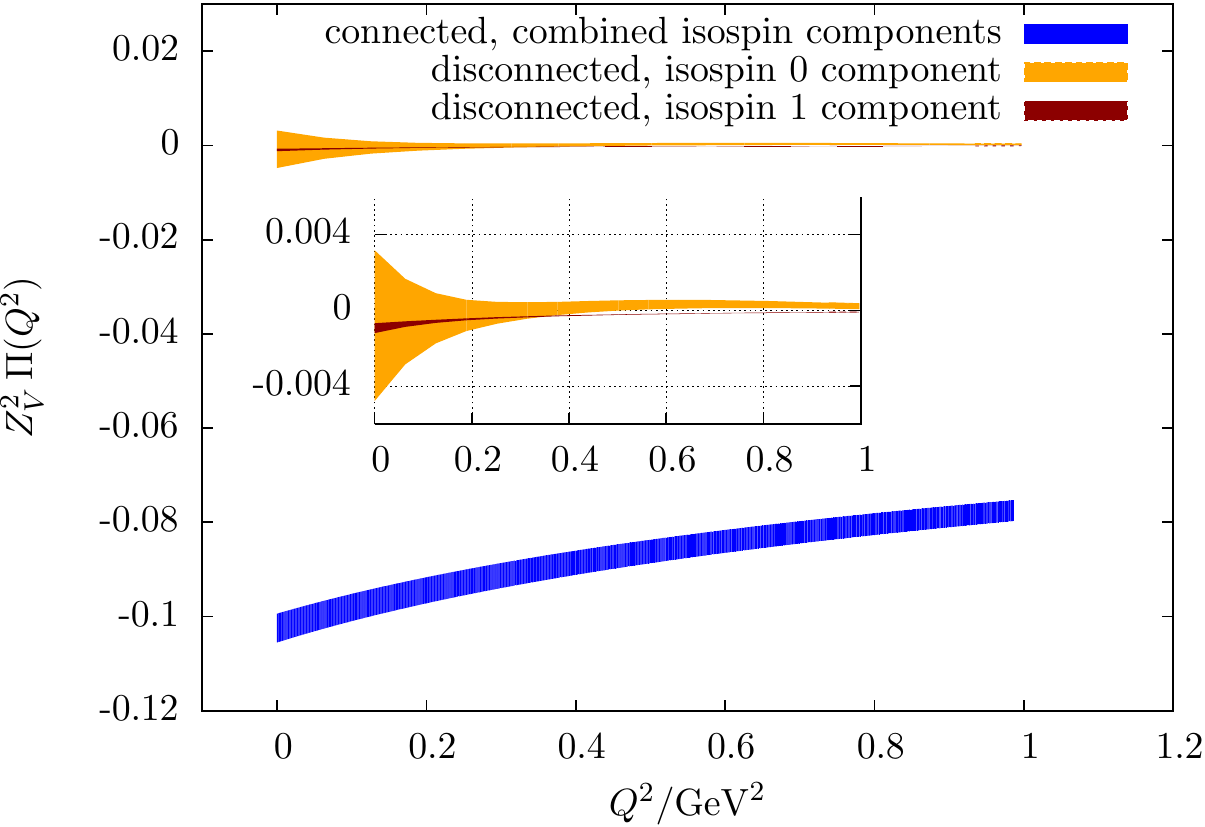}
\caption{Comparison of the light quark contributions to the unsubtracted hadronic vacuum polarisation function from quark-connected and disconnected
diagrams of the local current correlator. $Z_V$ has been
obtained from the ratio of the connected part of the conserved and local current-current correlators. The values have
been computed with the analytical continuation method described in~\cite{Feng:2013xsa} without correcting for finite-size effects.}
\label{fig:disconnected}
\end{figure}

Leaving out the quark-disconnected contributions is a systematic uncertainty we cannot completely quantify, yet. We
have started investigating their magnitude on the B55.32 ensemble mentioned already before. Using the local vector current we have detected a signal
for
the light quark part of the vacuum polarisation function when using 24 stochastic volume sources on 1548 configurations and 48 stochastic volume
sources on 4996 configurations. Employing the one-end trick~\cite{Boucaud:2008xu}, the isovector part
\begin{equation}
 \Pi^3_{\mu \nu}(x,y) = \langle J_\mu^3(x) J_\nu^3(y) \rangle
\end{equation}
with $J_\mu^3 = \frac{1}{2} \overline{\chi} \gamma_\mu \tau^3 \chi$
 is significantly different from zero.
However, this is a pure 
lattice artefact and will not contribute in the continuum limit. 
On the other hand, the more interesting isoscalar part 
\begin{equation}
 \Pi^0_{\mu \nu}(x,y) = \frac{1}{9}\langle J_\mu^0(x) J_\nu^0(y) \rangle
\end{equation}
with $J_\mu^0 = \frac{1}{2} \overline{\chi} \gamma_\mu \mathds{1} \chi$
is compatible with zero. 
The connected and disconnected pieces of the polarisation function for the light flavours are depicted 
in Fig.~\ref{fig:disconnected}.

A comparison of the values of $a_{l, \rm ud}^{\rm hvp}$ for all three leptons on the B55.32 ensemble with and
without incorporating the disconnected contributions is presented in Tab.~\ref{tab:disc}. Here, we have combined the connected pieces obtained from
the point-split current correlator with the isoscalar part of the disconnected contributions obtained from the local current correlator using the
renormalisation constant $Z_V$ determined from the ratio of the connected pieces of the conserved and the local vector current two-point functions. Therefore and because
we only have results for one ensemble, the numbers below can only give hints on the influence of the disconnected pieces. We observe the tendency that
for all
three leptons $a_{l, \rm ud}^{\rm hvp}$ decreases when incorporating the disconnected contributions as has been predicted
in~\cite{DellaMorte:2010aq}. However, this is statistically not significant. Furthermore, we find that the magnitude of the disconnected contributions is comparable to our current uncertainty.
Hence, it will be mandatory to compute them when aiming at more precise results. For the muon the value shifts by $\approx 3 \%$, which is also not
statistically significant at this stage, but is in accordance with
the upper bound of $4 - 5\%$ given in~\cite{Francis:2014hoa} as well as more recent high-statistics evaluations in \cite{Toth:2015lat,Blum:2015you}.

%\begin{table}[h]
%\begin{center}
%\begin{tabular}{|c|c c c|}
%\hline
% & & & \vspace{-0.40cm} \\
%%
% & $a_{\mathrm{e, ud}}^{\mathrm{hvp}}$ & $a_{\mathrm{\mu, ud}}^{\mathrm{hvp}}$& $a_{\mathrm{\tau, ud}}^{\mathrm{hvp}}$\\
%\hline \hline
% & & & \vspace{-0.40cm} \\
%without disc & $1.44(04)\cdot 10^{-12}$ & $5.42(14)\cdot 10^{-8}$ &   $1.27(03)\cdot 10^{-6}$ \\
%with disc & $1.39(07)\cdot 10^{-12}$  & $5.26(25) \cdot 10^{-8}$ &  $1.24(04)\cdot 10^{-6}$ \\
%\hline
%\end{tabular}
%\caption{\label{tab:disc} Comparison of light-quark contributions to $a_{l}^{\rm hvp}$ with and without disconnected pieces in the low-momentum
%region for the B55.32 ensemble. For all contributions the redefinition Eq.~(\ref{eq:redef}) and our standard analysis have been used.} 
%\end{center}
%\end{table}

\begin{table}[h]
\begin{center}
\begin{tabular}{|c|c c|}
\hline
% & & \vspace{-0.40cm} \\
& without disc & with disc \\
 \hline
 \hline
 $a_{\mathrm{e, ud}}^{\mathrm{hvp}}$  & $1.44(04)\cdot 10^{-12}$ & $1.39(07)\cdot 10^{-12}$ \\
 & & \vspace{-0.20cm} \\
 $a_{\mathrm{\mu, ud}}^{\mathrm{hvp}}$ & $5.42(14)\cdot 10^{-8}$ & $5.26(25) \cdot 10^{-8}$ \\
 & & \vspace{-0.20cm} \\
 $a_{\mathrm{\tau, ud}}^{\mathrm{hvp}}$ & $1.27(03)\cdot 10^{-6}$ & $1.24(04)\cdot 10^{-6}$ \\
% & & \vspace{-0.20cm} \\
\hline
\end{tabular}
\caption{\label{tab:disc} Comparison of light-quark contributions to $a_{l}^{\rm hvp}$ with and without disconnected pieces in the low-momentum
region for the B55.32 ensemble. For all contributions the redefinition Eq.~(\ref{eq:redef}) and our standard analysis have been used.} 
\end{center}
\end{table}

The disconnected heavy flavour contributions need to be considered as well.  We plan to check their size in future calculations.
The pure
charm quark contributions have been computed in perturbation theory and shown to be suppressed by a factor
$\left(\frac{q^2}{4~m_c^2}\right)^4$~\cite{Groote:2001py},
where $q^2$ is the relevant energy scale of the problem.

\subsection{Comparison with the phenomenological value}
Adding the quantified systematic uncertainties in quadrature we obtain as final result
\begin{equation}
 a_{\rm e}^{\rm hvp} = 1.782\,(64)(86)\cdot 10^{-12} \;.
\end{equation}
This can directly be compared with the phenomenological determination of~\cite{Nomura:2012sb}
\begin{equation}
 a_{\rm e}^{\rm hvp} = 1.866\,(10)\,(05)\cdot10^{-12}\;.
\end{equation}
They are fully compatible with each other although our lattice result
still is afflicted with larger errors.

\section{The $\tau$-lepton $(g-2)$}
\label{sec:tau}
The large mass of the tau lepton, $m_\tau \approx 1.8 \, {\rm GeV}$, implies a peak of 
the weight function in the expression for the LO hadronic 
contribution to its magnetic moment in Eq.~(\ref{eq:amudef}) at 
$Q_{\rm peak}^2= 0.745\, {\rm GeV}^2$. This is very different from the peak position of the electron weight function. 
Hence, the saturation of $a_{\rm \tau}^{\rm hvp}$ requires data from a different part of the subtracted vacuum polarisation function,
in particular, also the high-momentum piece of our fit function Eq.~(\ref{eq:pihigh}) is important here.

\subsection{Contribution from up an down quarks}
\begin{figure}[htbp]
 \centering
\includegraphics[width=0.48\textwidth]{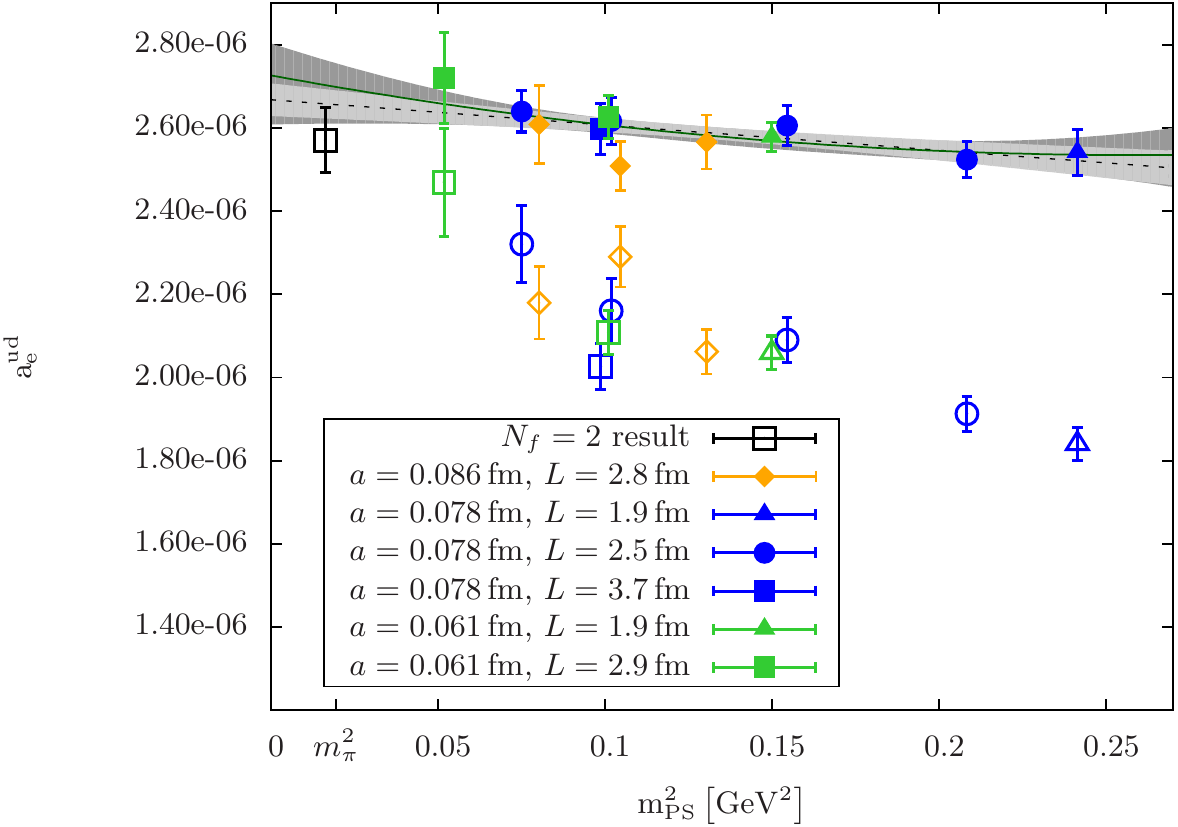}
\includegraphics[width=0.48\textwidth]{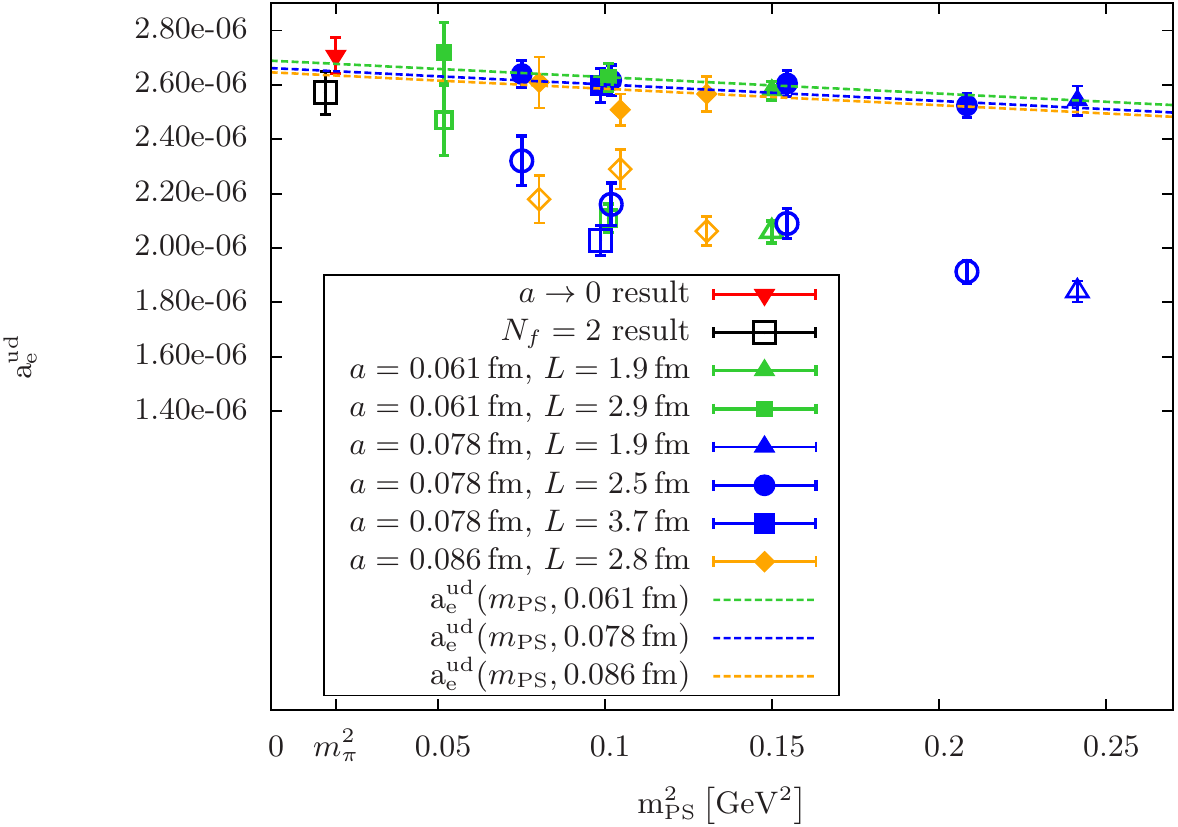}
\caption{
  Upper plot: light-quark contribution to $a_{\rm \tau}^{\rm hvp}$ 
  with filled symbols representing points obtained with Eq.~(\ref{eq:redef}), open symbols
  refer to those obtained with  Eq.~(\ref{eq:amudef}), i.~e.~$H=1$. We note that 
  the two-flavour result at the physical point has been computed with the standard
  definition. The light grey errorband belongs to the linear fit (dotted black line),
  whereas the dark grey errorband is attached to the quadratic fit (solid green line).
  Lower plot: combined chiral and continuum extrapolation taking into account leading order
  lattice artefacts.
}
\label{fig:ataulight_wphys}
\end{figure}

As for the electron, we start off by showing the contribution of the 
first-generation flavours to $a_{\rm \tau}^{\rm hvp}$ in the upper plot of Fig.~\ref{fig:ataulight_wphys}. 
The data show a qualitatively similar behaviour to those of the electron 
in Fig.~\ref{fig:aelight_wphys}. Their values differ, however, by six orders of magnitude. 
In particular, by comparing the upper and lower plot of Fig. \ref{fig:ataulight_wphys} 
we find that no significant lattice artefacts are present 
and that the data at unphysical pion masses obtained
with Eq.~(\ref{eq:redef}),  can 
be linearly extrapolated to the physical point. This demonstrates again 
that the method of including $\frac{H}{H_{\rm phys}}$ in the weight function is advantageous for the 
chiral extrapolation. The value extrapolated in this way and using all available lattice ensembles
agrees  with our calculation directly at the physical pion mass 
shown as the open square in Fig.~\ref{fig:ataulight_wphys}. 

\subsection{Adding the strange and the charm quark contributions}
\begin{figure}[htb]
 \centering
\includegraphics[width=0.48\textwidth]{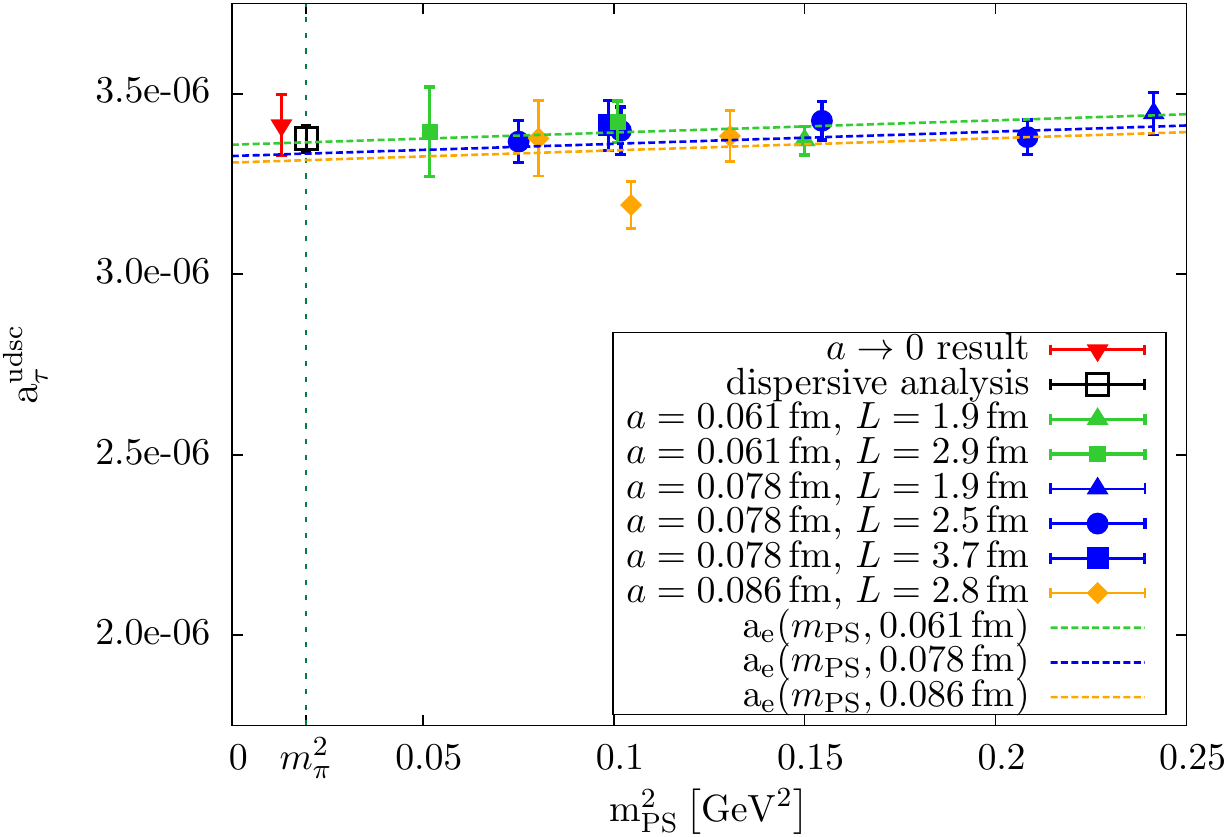}
\caption{Chiral and continuum extrapolation of the $N_f=2+1+1$ contribution to $a_{\rm \tau}^{\rm hvp}$. The inverted red triangle shows the value in
the continuum limit at the physical value of the pion mass. It has been displaced to the left to
facilitate the comparison with the dispersive result depicted as black square~\cite{Eidelman:2007sb}.}
\label{fig:atau_tot}
\end{figure}

\begin{figure}[htb]
\begin{minipage}{0.48\textwidth}
\includegraphics[width=0.95\textwidth]{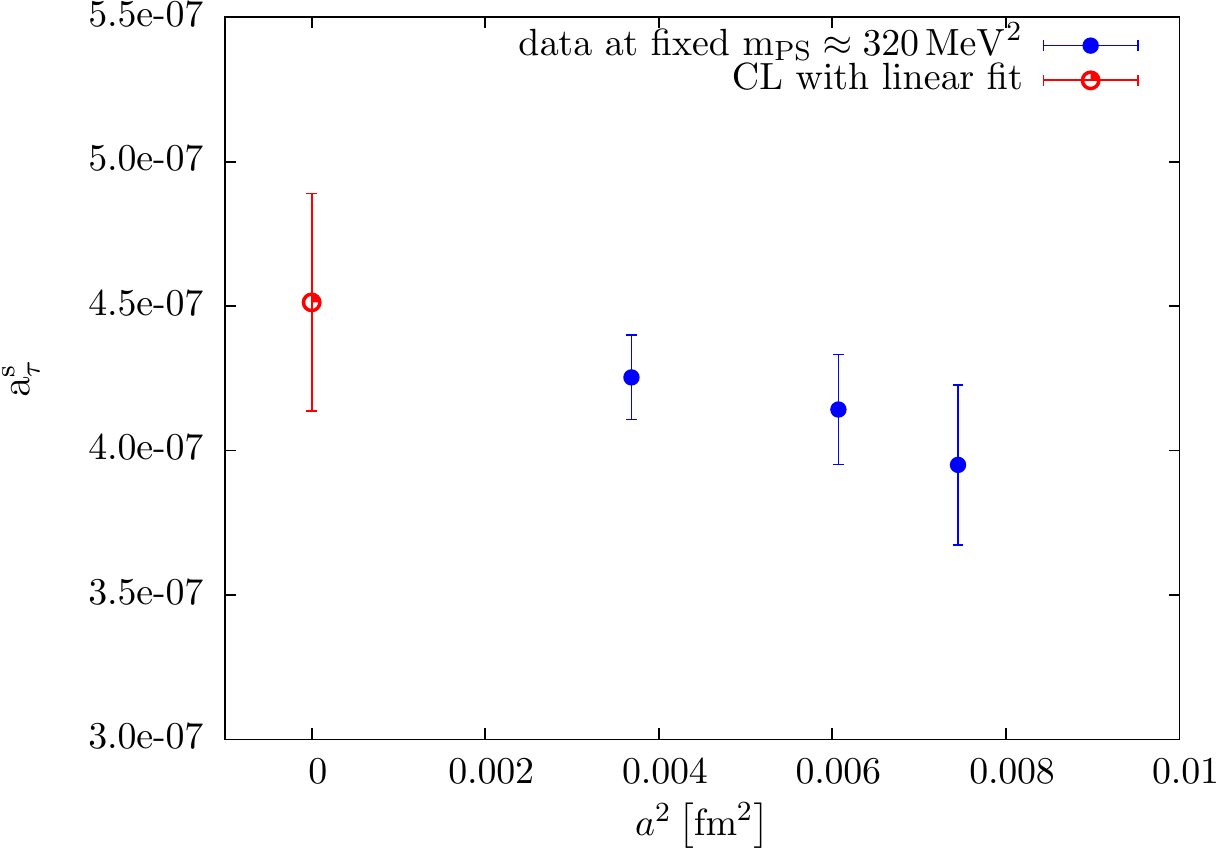}
\caption{Continuum limit of strange quark contribution to $a_{\rm \tau}^{\rm hvp}$ at approximately fixed pion mass.}
\label{fig:atau_strange}
\end{minipage}
\hspace{0.02\textwidth}
\begin{minipage}{0.48\textwidth}
\includegraphics[width=0.95\textwidth]{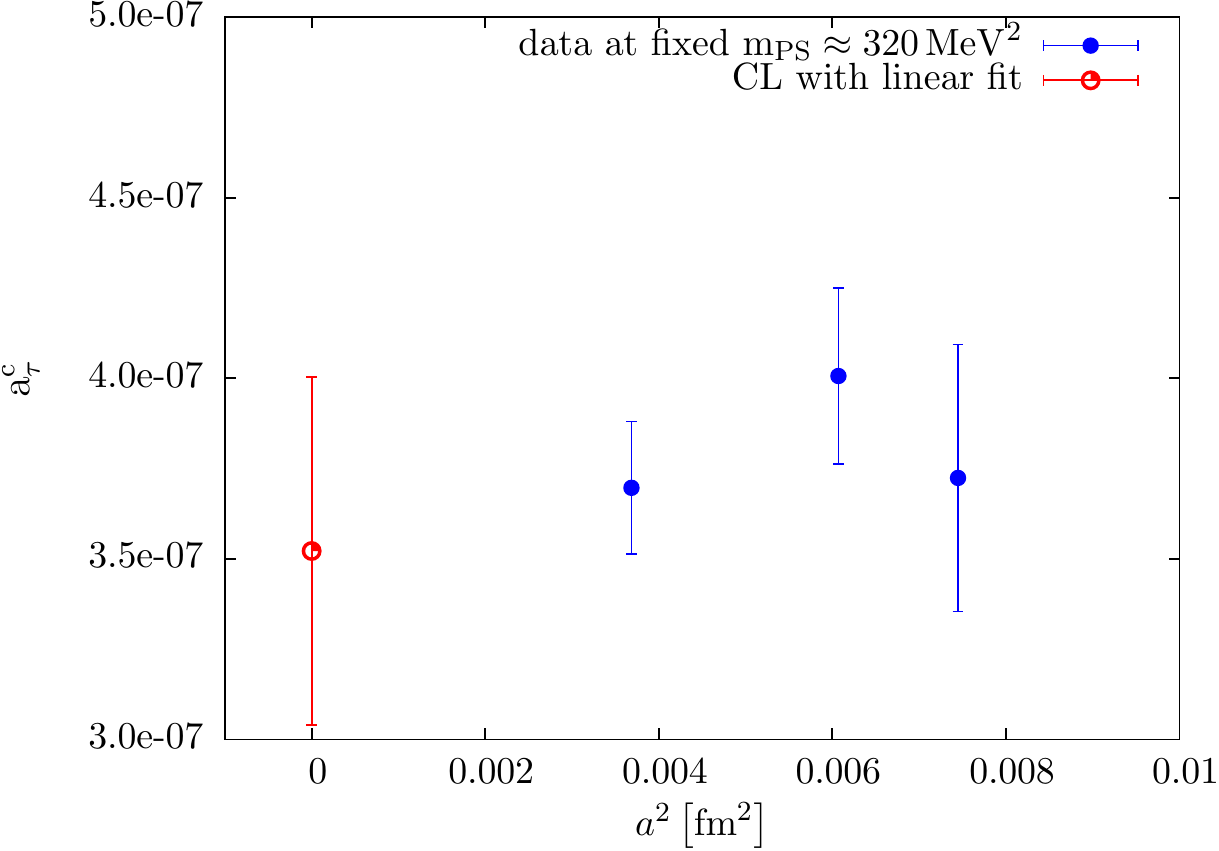}
\caption{Continuum limit of charm quark contribution to $a_{\rm \tau}^{\rm hvp}$ at approximately fixed pion mass.}
\label{fig:atau_charm}
\end{minipage}
\end{figure}

As for the electron, we perform the chiral and continuum extrapolation of the complete four-flavour result using a fit of the form given in \Eq{eq:fit}
and data for all lattice spacings, pion masses, and lattice volumes simultaneously.
It is shown in Fig.~\ref{fig:atau_tot}. Comparing this with
Fig.~\ref{fig:ae_tot}, we see that the lattice artefacts are much 
smaller than for the electron 
such that we would have obtained a compatible result when omitting the $a^2$ term in Eq.~(\ref{eq:fit}). 
As can be seen in
Figs.~\ref{fig:atau_strange} and \ref{fig:atau_charm}, for the tau lepton both, the strange and the charm
contribution do not show significant cut-off effects and hence, also for the total contribution $a^2$ effects are small. We nevertheless perform the
continuum extrapolation in order to use exactly the same analysis strategy as for the other leptons.

Our four-flavour result with only statistical uncertainty reads
\begin{equation}
 a_{\rm \tau}^{\rm hvp} = 3.41\,(8)\cdot10^{-6} \,.
\label{eq:atau_tot}
\end{equation}

\subsection{Systematic uncertainties}
We have investigated the same systematic uncertainties for our determination of $a_\tau^{\rm hvp}$ as for the case of the electron. The
influence of the disconnected contributions has already been discussed in the section of $a_{\rm e}^{\rm hvp}$.

\subsubsection{Finite size effects}
\label{sec:FSE_atau}
Restricting our data to the conditions $m_{\rm PS} L > 3.8$ and $m_\mathrm{PS} L > 4.5$ yields
\begin{align}
  a_{\rm \tau}^{\rm hvp}\left( m_\mathrm{PS} L > 3.8 \right) &= 3.40\,(09)\cdot 10^{-6} \; ,
  \label{eq:fse_a_tau_mpsL_gt_3.8}\\
  a_{\rm \tau}^{\rm hvp}\left( m_\mathrm{PS} L > 4.5 \right) &= 3.54\,(13)\cdot 10^{-6} \; .
  \label{eq:fse_a_tau_mpsL_gt_4.5}
\end{align}
This is compatible with the result in Eq.~(\ref{eq:atau_tot}). 
Comparing again the two ensembles at $m_{\rm PS} \approx 315\, {\rm MeV}$ which only differ in the extent of the lattices also indicates negligible finite size effects as shown in Tab.~\ref{tab:FSE_atau}.
Hence, we do not assign a FSE related systematic uncertainty.

 \begin{table}[htb]
\begin{center}
\begin{tabular}{|c c| c c|}
\hline
 & & & \vspace{-0.40cm} \\
Ensemble & $\left(\frac{L}{a}\right)^3 \times \frac{T}{a}$ & $a_{\mathrm{\tau, ud}}^{\mathrm{hvp}}$& $a_{\mathrm{\tau}}^{\mathrm{hvp}}$ \\
\hline \hline
 & & & \vspace{-0.40cm} \\
B35.32 & $32^3 \times 64$ & $2.62\,(06)\cdot 10^{-6}$ &   $3.40\,(07)\cdot 10^{-6}$ \\
B35.48 & $48^3 \times 96$ & $2.60\,(06) \cdot 10^{-6}$ &  $3.41\,(07)\cdot 10^{-6}$\\
\hline
\end{tabular}
\caption{\label{tab:FSE_atau} Comparison of light-quark contribution to $a_{\mathrm{\tau}}^{\rm hvp}$ 
and total $a_{\mathrm{\tau}}^{\mathrm{hvp}}$ from
ensembles of different volumes.} 
\end{center}
\end{table}

\subsubsection{Chiral extrapolation}
Restricting the analysed ensembles to those featuring pion masses $m_{\rm PS} < 400\,{\rm MeV}$, we get
\begin{equation}
 a_{\rm \tau}^{\rm hvp} = 3.45\,(09) \cdot 10^{-6}\; .
\end{equation}
This is again compatible with the value given in Eq.~(\ref{eq:atau_tot}). Hence, we do not assign a systematic uncertainty to the fact that ensembles with pion masses above $400 \; {\rm MeV}$ have been employed when extrapolating to the physical value of the pion mass.
 
\subsubsection{Vector meson fit ranges}
\begin{figure}[htbp]
 \centering
\includegraphics[width=0.42\textwidth]{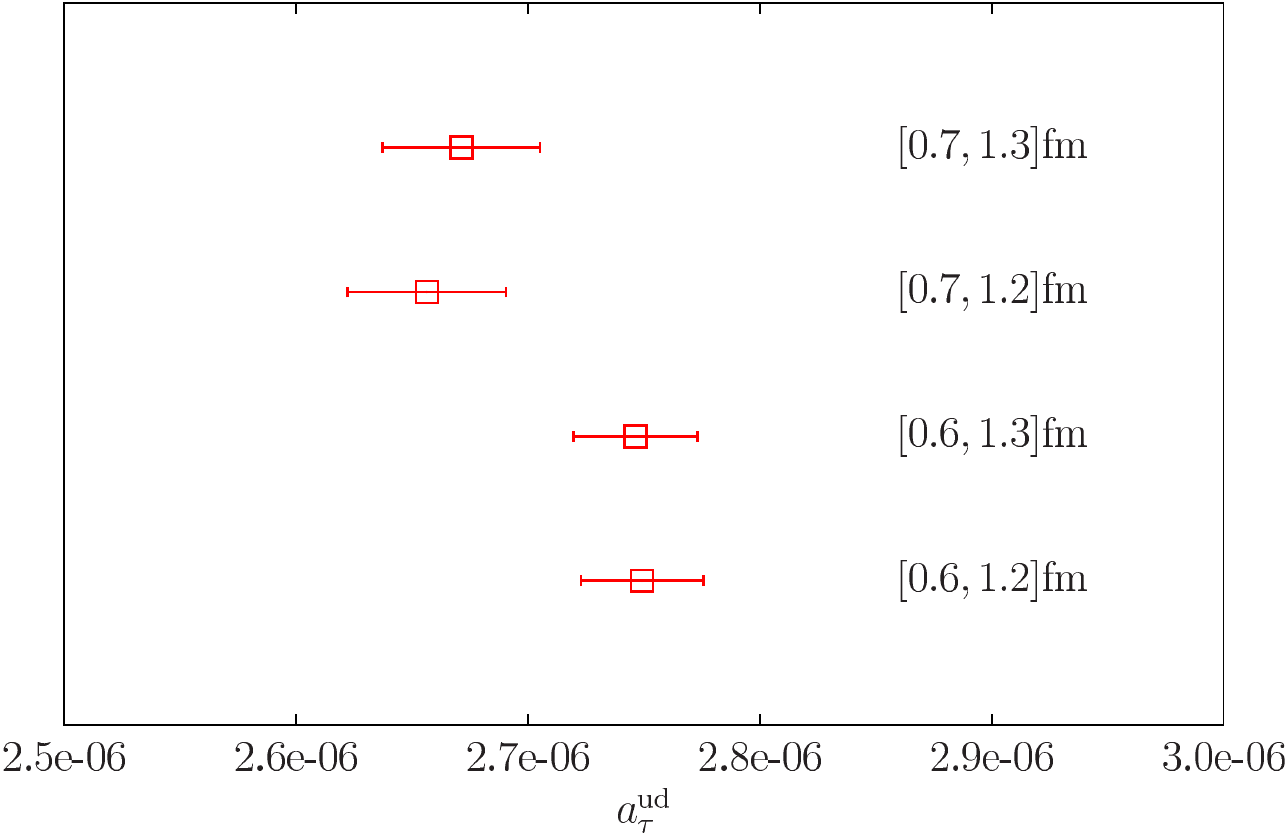}
\includegraphics[width=0.42\textwidth]{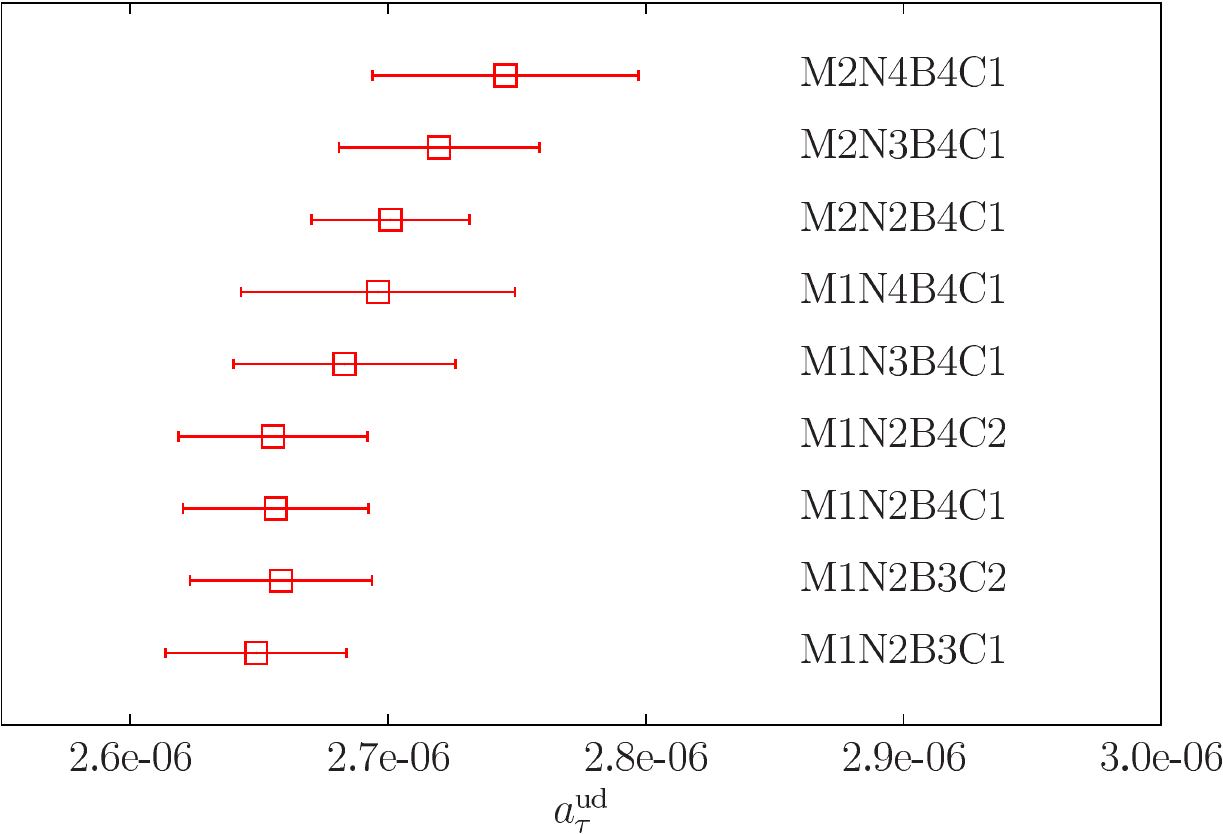}
\caption{Dependence of $a_{\rm \tau}^{\rm ud}$ on the fit range of the $\rho$-correlator (upper plot) and
on the values chosen for M, N, B, and C in the vacuum polarisation fit function (lower plot). The standard $\rho$-correlator fit
range is $[0.7\, {\rm fm},1.2\,{\rm fm}]$ and the standard fit function corresponds to M1N2B4C1.}
\label{fig:atau_fitrange_mnbc_light}
\end{figure}

The situation is similar to the case of the electron reported above. Only the excited state contamination in the $\rho$-correlator has to be
taken into account
as systematic uncertainty. In the upper plot of Fig.~\ref{fig:atau_fitrange_mnbc_light}
the dependence of the light quark contribution, $a_{\rm \tau}^{\rm ud}$,
on the fit range chosen to extract the spectral information from the $\rho$-correlator is depicted. 

Taking half the difference of the central values obtained for $[0.6\, {\rm fm},1.2\,{\rm fm}]$ and our standard fit range $[0.7\, {\rm fm},1.2\,{\rm fm}]$ results in an 
estimated 
systematic uncertainty of
\begin{equation}
 \Delta_{V} = 0.046 \cdot 10^{-6} \; .
\end{equation}

\subsubsection{Number of terms in MNBC fit function}
Due to the large $Q^2_{\rm peak}$ we have to take the whole vacuum polarisation function Eq.~(\ref{eq:pilowandhigh}) into account, including in
particular the high-momentum piece in Eq.~(\ref{eq:pihigh}). Thus, we have four different types of terms in the fit function that can have different
numbers of summands. We only find observable differences in the light quark sector. But also here the results from different fits
are all compatible as shown in the lower plot of Fig.~\ref{fig:atau_fitrange_mnbc_light}.  
Conservatively, we take
half the difference between the M2N3B4C1 and M1N2B4C1 fit and assign a systematic uncertainty of
\begin{equation}
 \Delta_{MNBC} = 0.032 \cdot 10^{-6} 
\end{equation}
to our choice of the fit function.

\subsection{Comparison with the phenomenological value}
Including the identified systematic uncertainties added in quadrature, our final four-flavour result reads
\begin{equation}
 a_{\rm \tau}^{\rm hvp} = 3.41\,(8)(6)\cdot10^{-6} \,.
\end{equation}
 This agrees with the one obtained by a dispersive analysis~\cite{Eidelman:2007sb}
\begin{equation}
 a_{\rm \tau}^{\rm hvp} = 3.38\,(4)\cdot10^{-6} \; .
\end{equation}
Compared to the electron, even better agreement between the lattice and the phenomenological result is observed for the $\tau$-lepton.
In this case, the uncertainty of our twisted mass LQCD calculation is only about twice the phenomenological one.

\section{Summary and Conclusions}
\label{sec:conc}
In this article we have presented the first four-flavour LQCD computation 
of the LO hadronic vacuum polarisation contributions to the anomalous 
magnetic moments of the electron and the $\tau$-lepton. Our results have been 
obtained with $N_f=2+1+1$ twisted mass fermions mostly at unphysically large 
pion masses but, at least for the light quark contribution, also directly at the 
physical point. We find that for both, the electron and the tau lepton, the 
chirally extrapolated values for the light quark contributions agree with 
the one at the physical point. 

For our data at unphysically large values of the pion mass we have investigated 
the systematic uncertainties of the method used to obtain our final results. In particular, we have addressed the effects of
non-zero lattice spacings, the finite volumes, the fit range for extracting the 
vector meson mass, and using different fit functions for the vacuum polarisation 
function. 
As an additional uncertainty we have investigated the disconnected contributions 
on one of our 4-flavor ensembles (B55.32)
by using the local vector current. This led to the first observation of a signal for the 
disconnected diagrams during our calculations, which, however, is compatible with zero within our current errors
and which we therefore have neglected. This will no longer be justified once the uncertainties of the connected
pieces are reduced and a full quantification of the quark-disconnected contribution will become significant.

Our final results are summarised in Tab.~\ref{tab:results_gm2} below and agree with the phenomenological determinations of the electron and tau
lepton 
magnetic moments which are also shown there.
This universal agreement across all three leptons and thus distinct weightings of the subtracted polarisation functions
  is elucidated by our findings in Ref. \cite{Burger:2015lqa}. There it was shown, that the subtracted vacuum polarisation function itself calculated
  with the methods used in this work and described in more detail in Ref. \cite{Burger:2013jya}, is compatible with the phenomenological
  result for $\Pi_R(Q^2)$ in the range $0 \le Q^2 \le \mathcal{O}\left( 10\, \mathrm{GeV}^2\right)$.

%%% This constitutes another evidence that our analysis, also employed for the muon and described in more detail in~\cite{Burger:2013jya}, is correct.  
\begin{table}[htb]
\begin{center}
%%%
%\begin{tabular}{c|l l l}
%  &  $a_{\mathrm{e}}^{\rm hvp}$ & $a_{\mathrm{\mu}}^{\mathrm{hvp}}$ & $a_{\mathrm{\tau}}^{\mathrm{hvp}}$ \\
%\hline \hline
% & & & \vspace{-0.40cm} \\
%this work & $1.782\,(64)(86)\cdot10^{-12}$ & $6.78\,(24)(16)\cdot 10^{-8}$ &   $3.41\,(8)(6)\cdot10^{-6}$ \\
%dispersive analyses & $1.866\,(10)\,(05)\cdot10^{-12}$~\cite{Nomura:2012sb} & $6.91\,(01)\,(05) \cdot 10^{-8}$~\cite{Jegerlehner:2011ti}& 
%$3.38\,(4)\cdot 10^{-6}$~\cite{Eidelman:2007sb}\\
%\end{tabular}
%%%
\begin{tabular}{|c|l l l|}
\hline
  & this work & dispersive analyses & \\
\hline
\hline
  $a_{\mathrm{e}}^{\rm hvp}$         & $1.782\,(64)(86)\cdot10^{-12}$ & $1.866\,(10)\,(05)\cdot10^{-12}$ & \cite{Nomura:2012sb} \\
  & & & \vspace{-0.20cm} \\
  $a_{\mathrm{\mu}}^{\mathrm{hvp}}$  & $6.78\,(24)(16)\cdot 10^{-8}$  & $6.91\,(01)\,(05) \cdot 10^{-8}$ & \cite{Jegerlehner:2011ti} \\
  & & & \vspace{-0.20cm} \\
  $a_{\mathrm{\tau}}^{\mathrm{hvp}}$ & $3.41\,(8)(6)\cdot10^{-6}$     & $3.38\,(4)\cdot 10^{-6}$         & \cite{Eidelman:2007sb} \\
\hline
\end{tabular}
\caption{\label{tab:results_gm2} Comparison of our first-principle values for $a_{\mathrm{e}}^{\rm hvp}$,
$a_{\mathrm{\mu}}^{\mathrm{hvp}}$, and
$a_{\mathrm{\tau}}^{\mathrm{hvp}}$ with phenomenological results. } 
\end{center}
\end{table}
As expected from the graph in the lower plot of Fig.~\ref{fig:saturation} the relative statistical uncertainties in all three
cases are similar. For the electron the systematic uncertainty already exceeds the statistical one.

As in the case of the muon, also for the electron and tau lepton anomalous magnetic 
moments the errors of our calculations are still larger than those 
from the dispersive analyses quoted above.   
However, it can be expected that with future lattice QCD calculations at the physical value 
of the pion mass, increased statistics and an even better control over systematic 
uncertainties the phenomenological error can be matched, if not even beaten, especially for the $\tau$-lepton.

\begin{acknowledgements}
We thank the European Twisted Mass Collaboration (ETMC) for generating the gauge field ensembles used in this work. Special thanks goes to the authors
of~\cite{Michael:2013gka} who generously granted us access to their data for the disconnected contributions of the
local vector current correlators.
This work has been supported in part by the DFG Corroborative
Research Center SFB/TR9.
G.P.~gratefully acknowledges the support of the German Academic National Foundation (Studienstiftung des deutschen Volkes e.V.) and of the
DFG-funded Graduate School GK 1504.
The numerical computations have been performed on the
{\it SGI system HLRN-II} and the
{\it Cray XC30 system HLRN-III} at the {HLRN Supercomputing Service Berlin-Hannover},  FZJ/GCS, BG/P, and BG/Q at FZ-J\"ulich.
\end{acknowledgements}

\bibliographystyle{spphys}       % APS-like style for physics

\bibliography{gm2}

\end{document}